\documentclass[a4paper,12pt]{article}
\usepackage{booktabs}
\usepackage{multirow}
\usepackage{rotating}
\usepackage{color}
\usepackage{xcolor}

\usepackage{soul}

\usepackage{mathtools}  
\mathtoolsset{showonlyrefs}  

\usepackage{graphicx}
\usepackage{subfigure}

\begin{document}

\title{Specific adsorption of phosphate species on Ag (111) and Ag (100) electrodes and their effect at low overpotentials of the hydrogen evolution reaction}
\author{Claudia B. Salim Rosales$^{(1)}$, Mariana I. Rojas$^{(2)}$ and Luc\'ia B. Avalle$^{(1)}$
\thanks{Corrresponding author. e-mail:avalle@famaf.unc.edu.ar}\\
(1) IFEG, CONICET, FaMAF, UNC, C\'ordoba, Argentina.\\
(2) INFIQC (CONICET), Fac. de Ciencias Qu\'imicas (U.N.C.), \\ 
C\'ordoba, Argentina.}

\date{}
\maketitle

\begin{abstract}
Investigation of phosphate species adsorption/desorption
processes was performed on Ag(100) and Ag(111) electrodes
in H$_{3}$PO$_{4}$, KH$_{2}$PO$_{4}$ and K$_{3}$PO$_{4}$ solutions
by Current-Potential ($j-V$) profiles and
Electrochemical Impedance Spectroscopy ($EIS$).
We used the equivalent circuit 
method to fit the impedance spectra.
Different electrical equivalent circuit ($EEC$) were employed depending on the potential region that was analyzed.
For potentials more negative than the onset of the hydrogen evolution reaction ($her$), a charge transfer resistance (R$_{ct}$) in parallel to the $(RC)$ branches was included.
Peaks from $j-V$ profiles were integrated to estimate surface coverage.
A reversible process was observed for Ag(hkl)/KH$_{2}$PO$_{4}$ systems,
where a value of 0.07 ML was obtained. For Ag(111)/H$_{3}$PO$_{4}$, a coverage of about 0.024 ML was calculated from anodic/cathodic $j-V$ profiles, whereas for Ag(hkl)/K$_{3}$PO$_{4}$ systems different values were obtained from integration of anodic/cathodic peaks due to highly irreversible processes were observed. In the case of Ag(hkl)/K$_{3}$PO$_{4}$, the capacitance (C$_({\phi})$) plots are well differentiated for the two faces, and co-adsorption of OH$^{-}$ was evaluated from resistance parameters.
Characteristic face-specific relaxation times are obtained 
for each electrode.
In addition, it was found that the onset potential of $her$ for Ag(111) at pH=1.60 was about
100 mV more negative compare to Ag(100).
\end{abstract}

keyword: phosphates, adsorption, onset potential of $her$, relaxation time,
single crystals, silver.

\section{Introduction}
The development of efficient devices for energy conversion
strongly depends on the
knowledge of the reactivity in
the electrode/electrolyte interfaces
for reactions such as
the hydrogen evolution reaction ($her$),
and hydrogen oxidation reaction ($hor$)
\cite{lamoureux2019, dubouis2019}.
New studies focusing on the role of
buffer species
in the $her$ performance of
different materials
have shown how the nature of the anion impacts
the $her$ kinetics \cite{sarabia2019}.
This demonstrates the important effect of phosphate species
and other oxoanions \cite{ruderman2013}
on the reaction
mechanisms when
they are specificaly adsorbed
at the interface.

We have recently studied
the strength of interaction of phosphate species on Ag(hkl)
surfaces by Cyclic Voltammetry, Current Transients and
vdW-DFT calculations \cite{salim2017}. In those studies, it
has been demonstrated that the adsorption of phosphate species on
silver electrodes is highly sensitive to the surface structure,
and it was found that the
stability of the
adlayer strongly increased
with hydrogen loss
in the phosphate molecules,
being more significant for Ag(100)
than for Ag(111) surface \cite{salim2017}.
Kolovos-Vellianitis et al. have
studied the abstraction of D on silver single crystal surfaces by gaseous H atoms in ultra-high-vacuum conditions by thermal desorption spectroscopy \cite{kolovos2004}. Temperatures of 150 K and 160 K were reported for HD abstraction of Ag(100) and Ag(111), respectively.
However, Kuo et al. have used DFT calculations in vacuum and have reported
Ag-H binding energies of -0.35 eV and -0.12 eV on Ag(100) and Ag(111), respectively \cite{kuo2019}. Although our systems are at electrochemical conditions, if Ag-H is still the main interaction, we could expect during negative overpotential scan that protons adsorption and hydrogen evolution could appear first on Ag(100) and then on Ag(111).

It has been demonstrated that the conformational geometry of the adsorbed phosphate
species on Ag(hkl) surfaces
is potential-dependent, where at -1.2 V vs. SCE \cite{salim2017, niaura2001}
a monodentate coordination between an oxygen atom
from the phosphate molecule
and a silver atom on the electrode surface is stablished.
This coordination changes from mono to bi, and even to tri-dentate,
when changing the electrode potential.
Therefore, they follow the mentioned trend (mono, bi and tri)
as the applied potential turns towards more positive values.
Also, the ordered submonolayers of phosphate species have the symmetry of the substrate, 
C$_{3v}$ for Ag(111) and C$_{4v}$ for Ag(100), respectively \cite{salim2017}.

Sundararaman et al. \cite{doi:10.1063/1.5024219} have calculated
by first-principles methods the variation of C($\phi$)
as a fuction of potential, reproducing the experimental curves
observed for Ag(100) electrodes in KPF$_{6}$ electrolyte \cite{valette198237}.
Other theoretical approaches are also being currently applied
for the microscopic explanation of the differential
capacitance of the double layer \cite{may2019125}.

In the present work, we study the (100) and (111) crystallographic orientations of silver in phosphate electrolytes by
electrochemical techniques.
The adsorption of phosphate
species together with co-adsorption of other species was studied by $j-V$
profiles, and results were compared to those from $EIS$.

From the impedance spectra, the kinetic parameters
of the adsorption have been evaluated
and correlated with
the adsorption specificity
on the two low-index
planes of silver.
Characteristic curves of capacitance (C($\phi$))
as a function of potential ($\phi$)
were obtained.
The calculated relaxation times
were associated with
changes at the electrode surfaces
induced by the presence of phosphate species.
We also report the first results of
the onset potential of $her$ of silver electrodes in phosphate-containing
electrolytes.

\section{Experimental measurements}
\subsection{Materials}
A three-electrode system was used for
the electrochemical measurements.
The working electrodes were Ag(hkl) single crystals provided by Mateck
(orientation accuracy $<$ $1\,^{\circ}\mathrm{}$).
High purity polycrystalline annealed Pt wire (99.999\%)
was used as a counter electrode. The
reference electrode was a saturated
calomel electrode (SCE).

Pretreatment of the crystals was performed as
described in Ref. \cite{avalle2011}.
Here we outline the new steps that were not reported previously.

After a chemical polishing, \cite{avalle2011}
the electrodes were treated in high-purity nitrogen
gas at 600$\,^{\circ}\mathrm{}$C during 5 min, followed by
400$\,^{\circ}\mathrm{}$C during 15 min and cooling down
to room temperature in a continuous gas flow, and then by the 
electrode transfer to the electrochemical cell
with a drop of ultrapure water. The contact between the electrode 
surface and the solution was performed by the hanging meniscus 
method under potential control.

As a first step, the results obtained in 0.01 M KClO$_{4}$ by $CV$ 
and $EIS$ were used to test the cleanliness of the surfaces. 
The measurements to determine the 
potential of zero charge ($pzc$) 
of the electrode surfaces 
were carried out in 0.01 M KClO$_{4}$.
This concentration is within the range where 
a Parsons and Zobel plot is linear \cite{santos2010}. 
Thus, this is an indication that 
specific adsorption from electrolyte does not occur.

Chemicals were from Aldrich (analytical grade, purity $>$ 99.9\%). 
The KClO$_{4}$ and KH$_{2}$PO$_{4}$ salts were recrystallized twice. 
The solutions were comprised of 0.01 M of 
the following electrolytes: KClO$_{4}$, 
H$_{3}$PO$_{4}$, KH$_{2}$PO$_{4}$ and K$_{3}$PO$_{4}$.

Oxygen-free solutions were obtained by continuously 
purging the electrochemical cell with nitrogen gas 
(99.999\% purity). 
Millipore System was used for the preparation 
of all solutions and also for glassware cleaning.

The reproducibility of the results was determined by repeating 
the measurements from a fresh initial electrode surface. 
By this method, voltammetric scans were recorded for freshly 
annealed surfaces, applying no more than 2.5 cycles, thus avoiding 
repetitive cycling that increases surface roughness. It should be 
noted that in this procedure, the electrode undergoes a greater 
number of heating cycles. 

\subsection{Current-potential ($j-V$) profiles}
The electrochemical analyzer was an AUTOLAB-PGSTAT302N. 
The $j-V$ profiles were recorded at a scan rate 
of 0.05 Vs$^{-1}$.

Measurements to determine the effect of phosphate species  
in the onset potential of $her$ were started at -0.20 V 
towards negative potential limits. In addition, voltammograms 
starting at a negative potential up to different 
positive limits were also recorded.
The baseline correction of the $j-V$ peaks was carried out 
using a polynomial function. The charge values were then 
obtained by integration as implemented in fityk free software 
(see Supporting Information ($SI$)).

\subsection{Electrochemical Impedance Spectroscopy (EIS)}
Measurements of electrochemical impedance spectra 
of silver electrodes in the frequency range 5 kHz - 0.1 Hz have 
been performed with a FRA32 M module from Autolab, and driven by
Nova 2.1 software (Metrohm). Causality was
determined using a Kramers-Kronig test included in the software, 
and linearity was checked by measuring
spectra at amplitudes 5-15 mV (rms).

To check the $EIS$ procedure, a dummy cell containing 
a combination of resistive and capacitive elements 
(connected in series and in parallel) 
forming different circuits 
was used to run the impedance spectra in the whole frequency range. 
The impedance values obtained from this run 
were fitted using the 
analysis procedure from Nova, with errors less than 1\% for each electrical element.

The experimental
admittance and dielectric complex permittivity 
spectra were as follows: $\epsilon$ = $\frac{1}{i\omega Z}$ (where i=$\sqrt{-1}$)
were used to plot the data, and 
the equivalent circuit method was employed 
to fit a physical model representing the 
electrode/electrolyte interface \cite{pajkossy201753, lagonotte2019256, ciucci2019132}. 
However, parameters of the
electric equivalent circuits ($EEC$) 
were fitted to actually measured
impedance, rather than to $\epsilon$($\omega$).

Characteristic relaxation times were obtained from the $EEC$s
by multiplying the R$_{}$ and C$_{}$ elements connected in
series between them, i.e., (R$_{1}$C$_{1}$) and (R$_{2}$C$_{2}$) branches.
The best fittings were used as an approach to measure the dynamical
structure of the inteface. Capacitance values are normalized
in relation to the geometrical area of the electrode.

As the equivalent circuit method has the disadvantage
that more than one equivalent circuit can be fitted to
the same experimental data (resulting in different
values for the parameters), providing an ambiguous
physical model, we have focused our attention on the experimental setup.
Therefore, Ag(hkl)/electrolyte interfaces were obtained
under highly controled conditions (see section 2.1),
where the silver oxide formation was avoided and highly
purified salts and solvent were used to prepare the
electrolyte.
Thus, this experimental procedure allows for reducing
the number of parameters that must be introduced, in
series or parallel connections, in the equivalent
circuit analog. As a consequence, the physical model
asociated to the electrochemical interface is also
simplified \cite{harrington2020}.

\section{Results}
Fig. 1 shows the $EECs$
used to represent the different interfaces.

\begin{figure}
\centering
  \includegraphics[height=0.75cm]{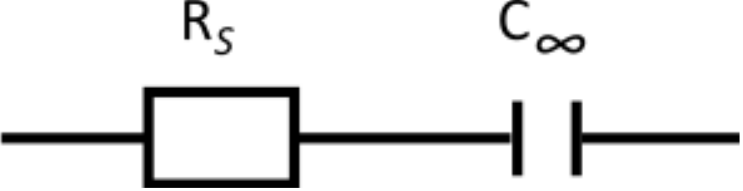}
\vspace{0.2cm}
{(a)}
\vspace{0.5cm}
  \includegraphics[height=2.00cm]{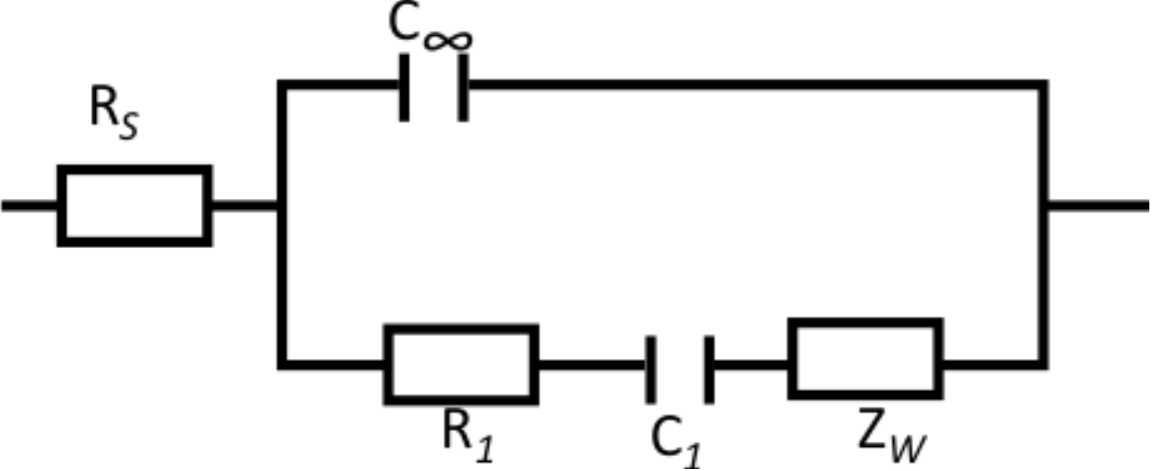}
{(b)}\\
\vspace{0.2cm}
\vspace{0.5cm}
  \includegraphics[height=2.00cm]{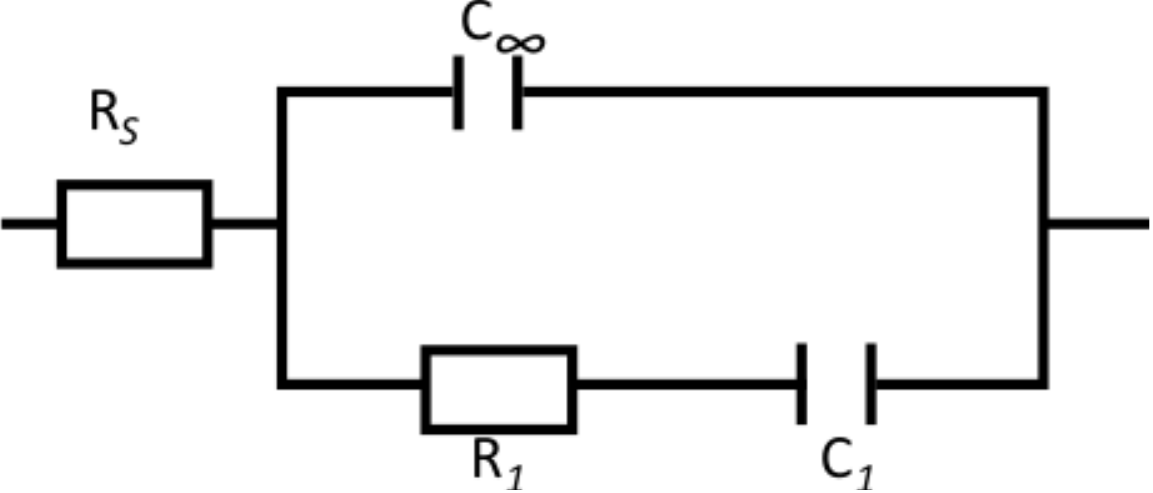}
\vspace{0.2cm}
{(c)}
\vspace{0.5cm}
  \includegraphics[height=2.50cm]{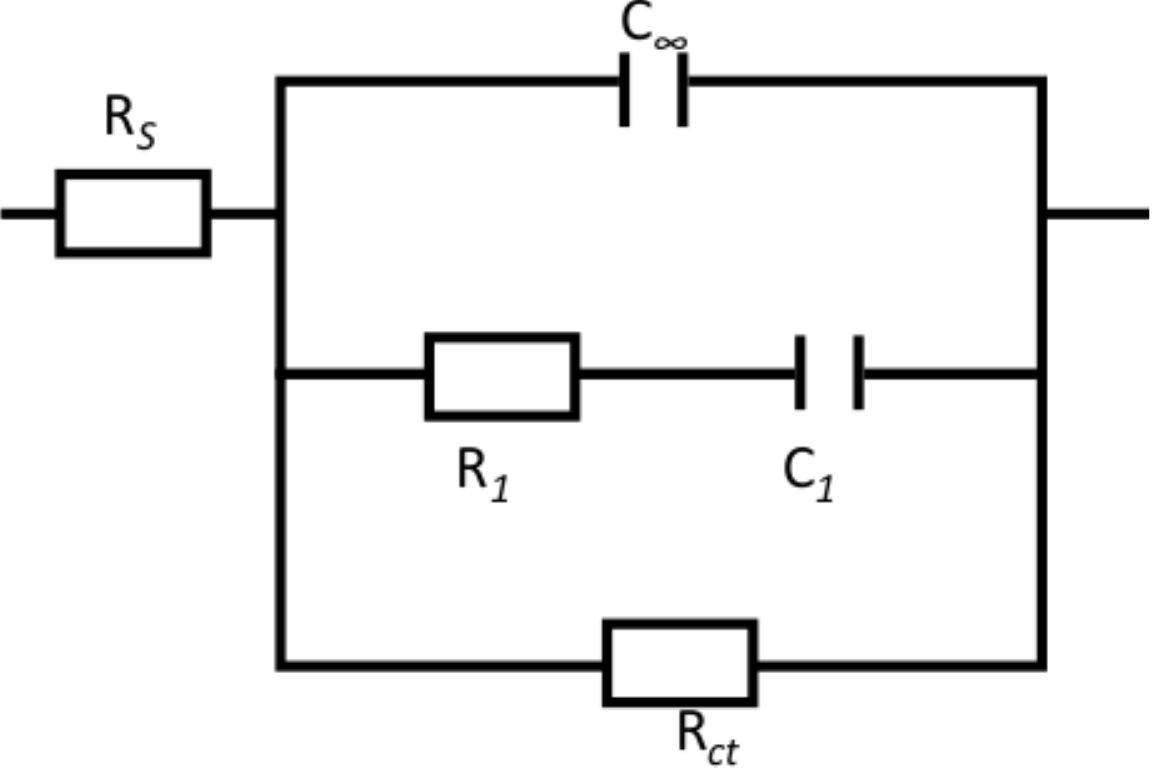}
\vspace{0.2cm}
{(d)}\\
\vspace{0.5cm}
  \includegraphics[height=3.00cm]{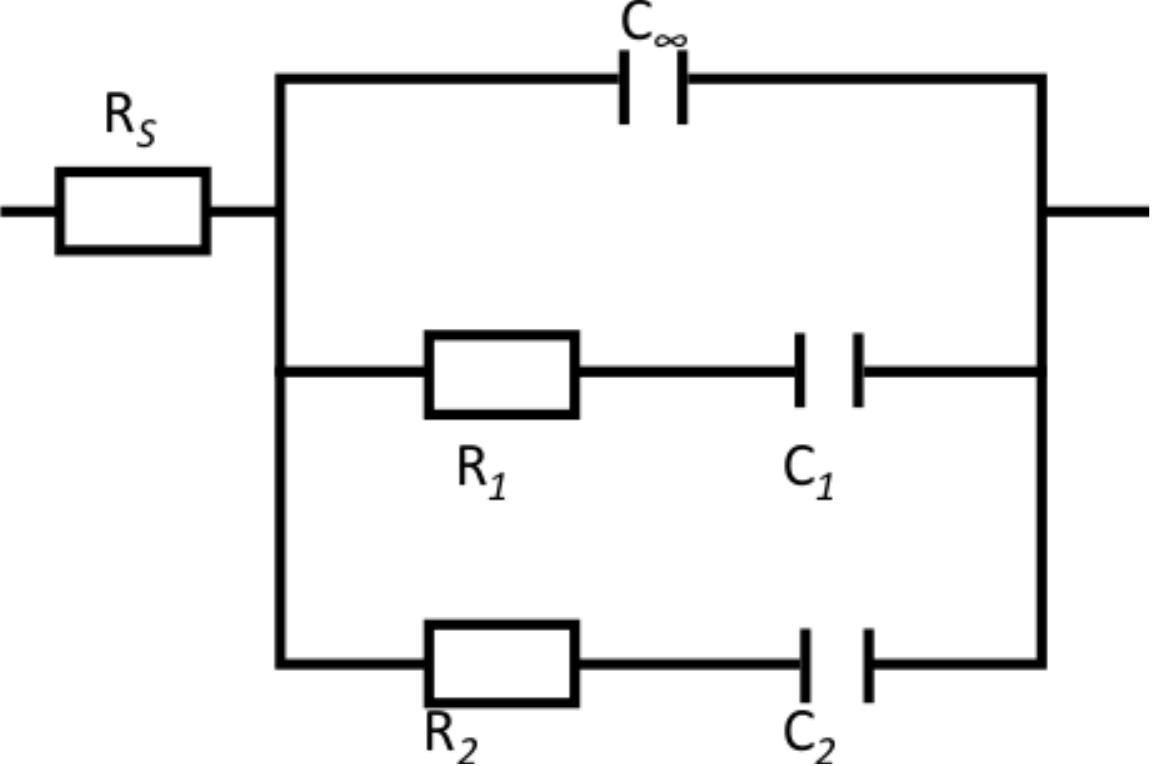}
{(e)}
\vspace{0.2cm}
  \includegraphics[height=3.50cm]{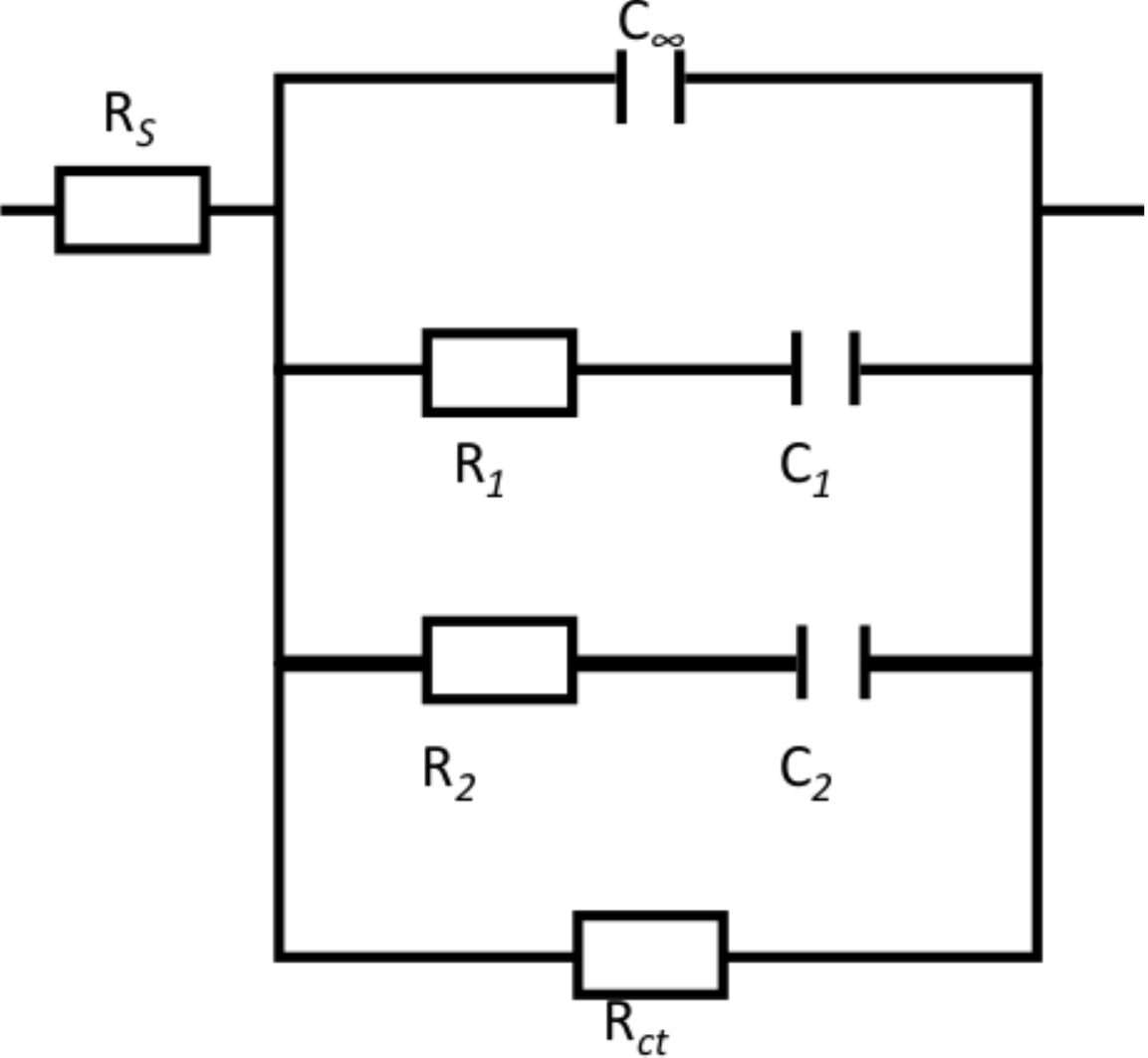}
{(f)}\\
\vspace{0.2cm}
  \caption{Electrical Equivalent Circuits ($EEC$) used for modeling surface impedance. 
(a) Fast processes. (b) Slow kinetic and diffusional processes. (c) Slow kinetic processes. C$\infty$: high frequency capacitance. R$_{sol}$: electrolyte resistance. R$_{1}$: adsorption resistance. C$_{1}$: adsorption capacitance. 
(d) Similar to (c) with addition of a R$_{ct}$ that represents the charge transfer resistance in the potential region where $her$ takes place. (e)  Similar to (c) with addition of a second branch, R$_{2}$ and C$_{2}$. These two branches take into account the co-occurrence of two processes. (f) Similar to (e) with addition of a R$_{ct}$ that represents the charge transfer resistance in the potential region where $her$ takes place.}
\end{figure}

The simple model represented by the $EEC$ of Fig. 1 (a) was run to obtain the parameters to be
used as a first guide for further improvements on the fitting
procedure.
The $EEC$ of Fig. 1 (b), which considers diffusion of
adsorbing ions from/to the interface, was also tested.

The $EEC$ of Fig. 1 (c) is formed by a series combination of
one (R$_{1}$C$_{1}$) branch associated with surface processes.
A charge transfer resistance (R$_{ct}$) is added in the $EEC$s of Fig. 1 (d)
and (f) to fit the experimental data in the potential region where the $her$ takes place.

Fig. 2a shows complex permittivity plots of Ag(hkl) electrodes in 0.01 M H$_{3}$PO$_{4}$. 
The $EIS$ spectra of Ag(111) (left) are shown for -0.20 V (black), 
-0.40 V (red), and -0.55 V (blue);
and the $EIS$ spectra of Ag(100) (right) corresponding to 
-0.20 V (black), -0.40 V (red), and -0.60 V (blue).

\begin{figure}
\centering
  \includegraphics[height=3.5cm]{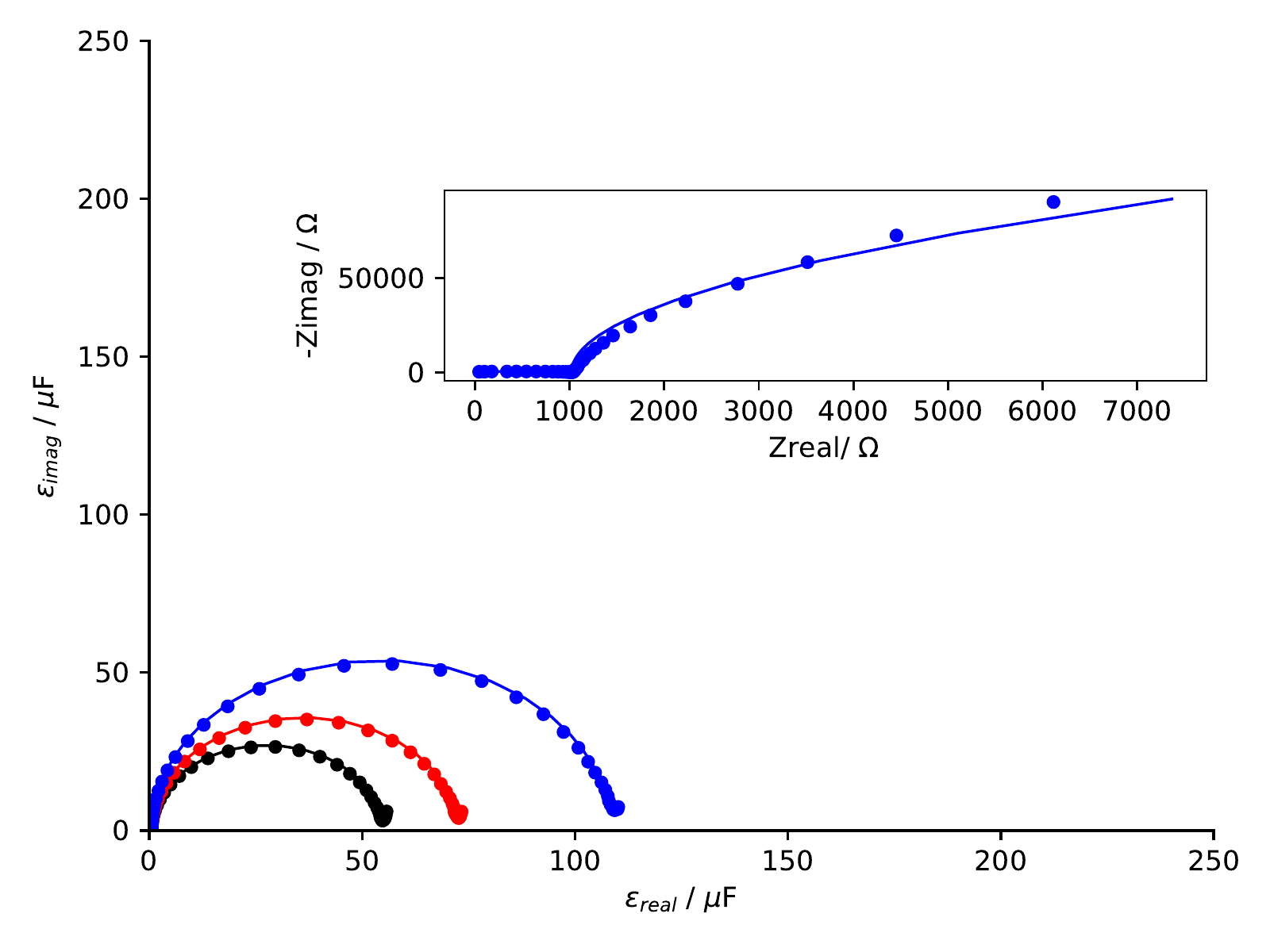}
  \includegraphics[height=3.5cm]{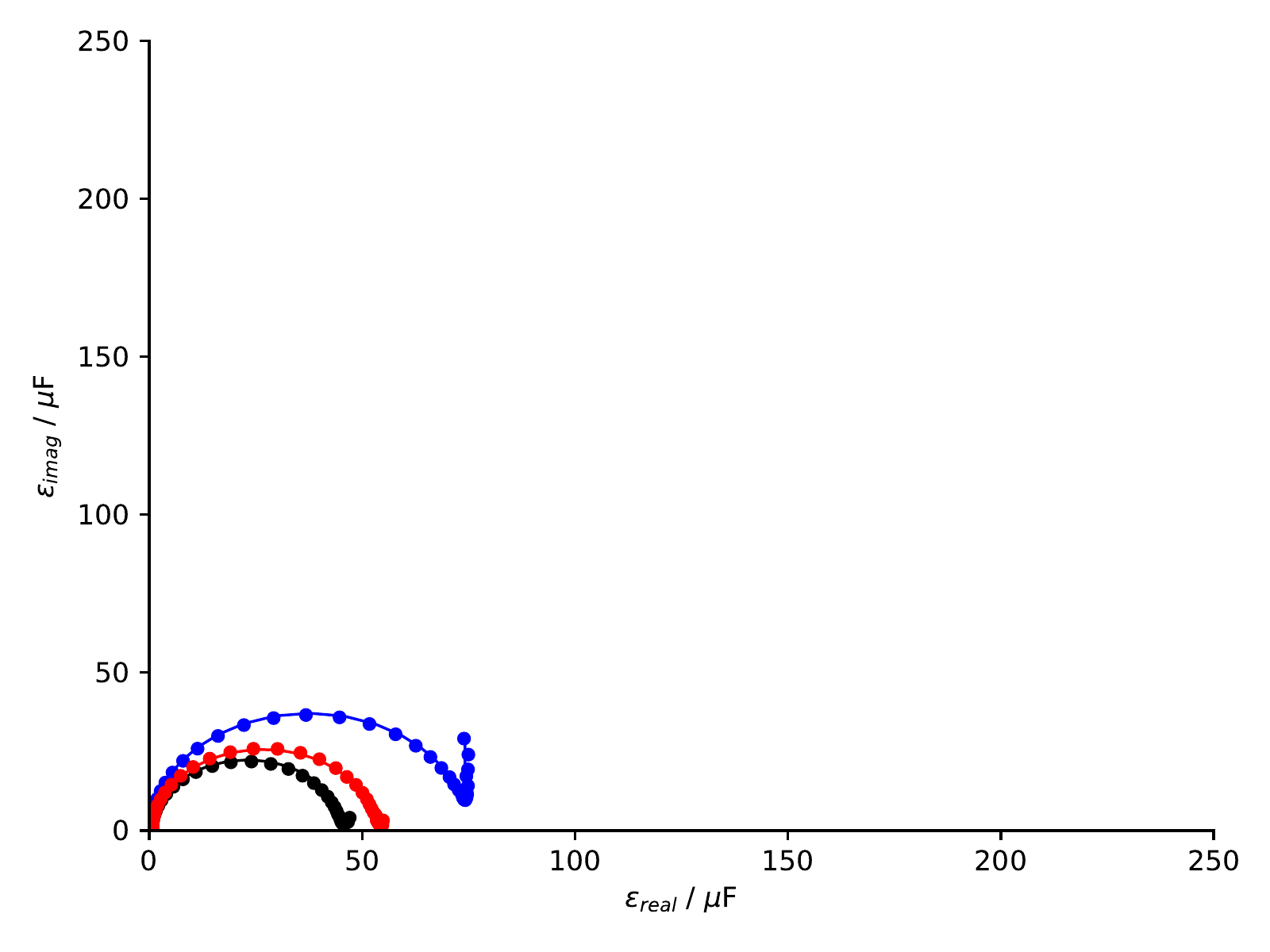}\\
\vspace{0.2cm}
{(a)}\\
  \includegraphics[height=3.5cm]{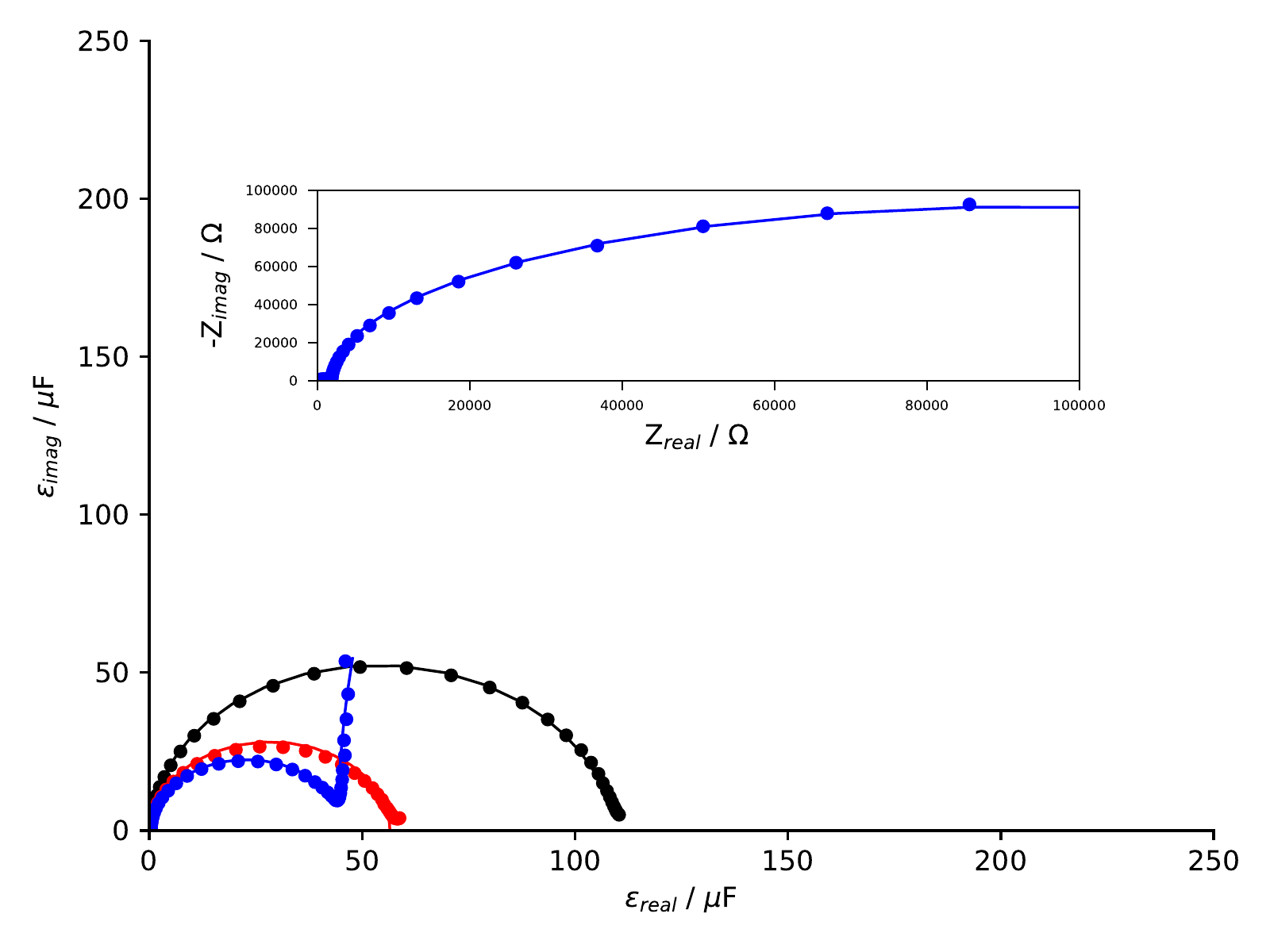}
  \includegraphics[height=3.5cm]{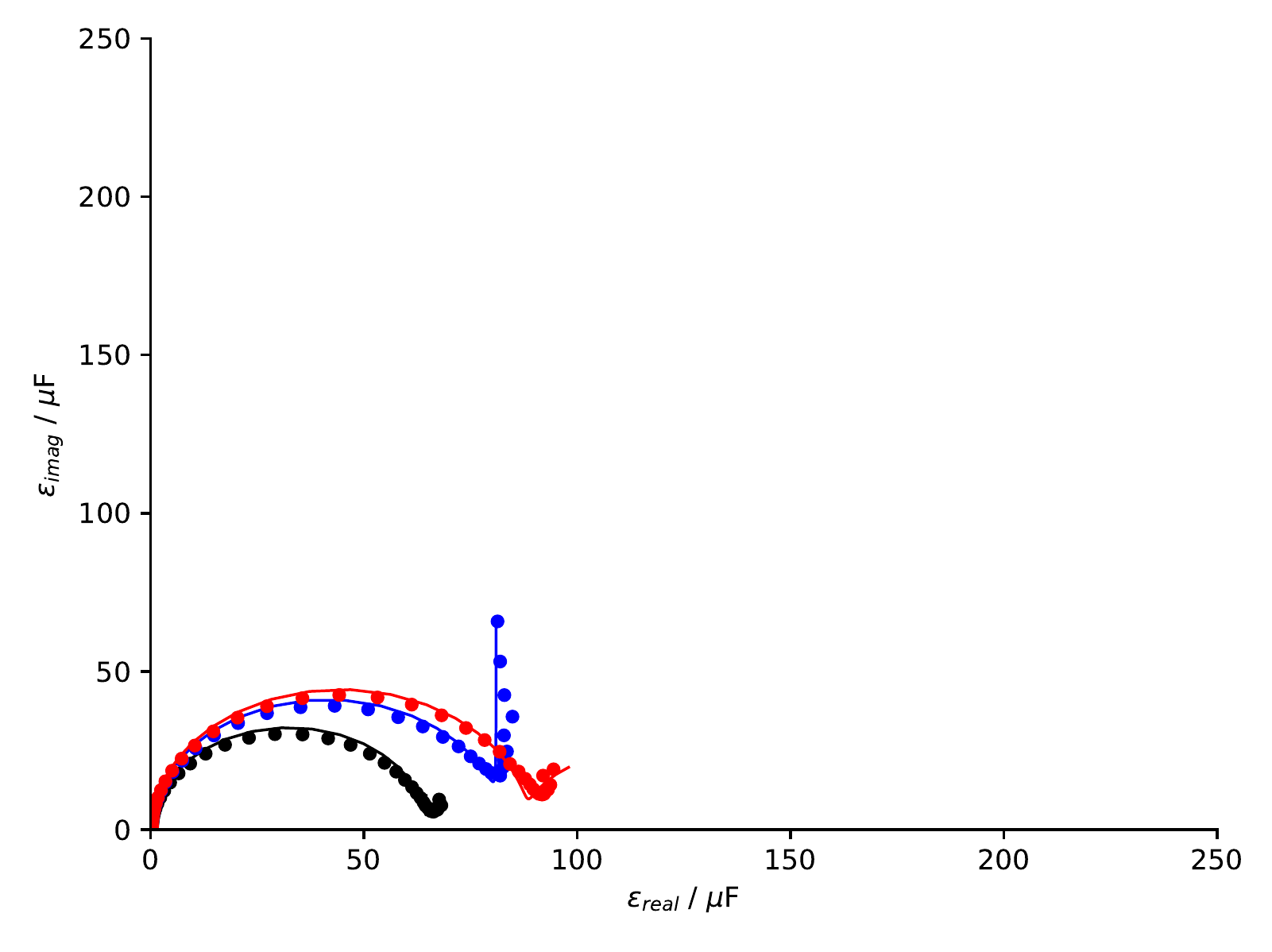}\\
\vspace{0.2cm}
{(b)}\\
  \includegraphics[height=3.5cm]{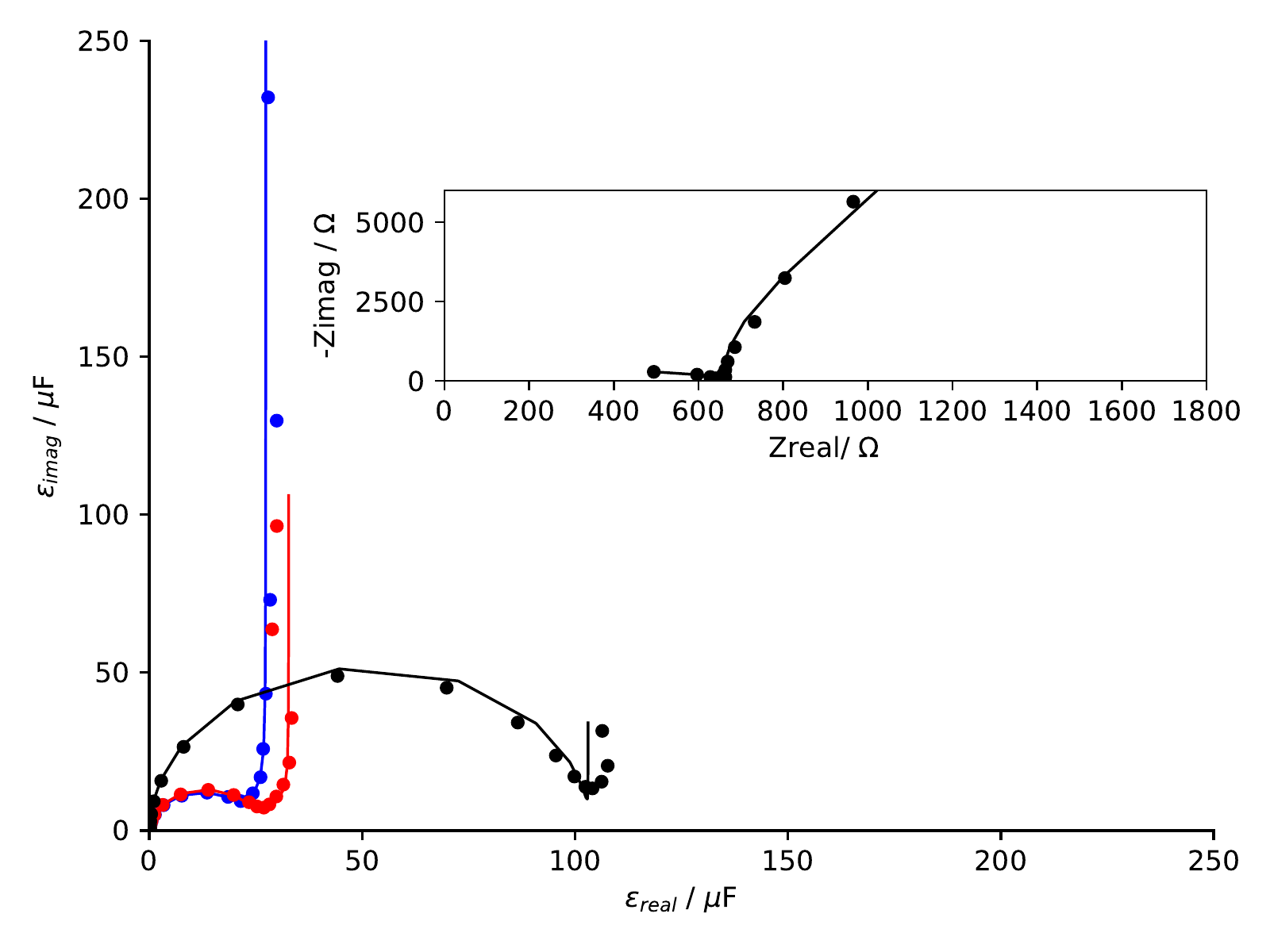}
  \includegraphics[height=3.5cm]{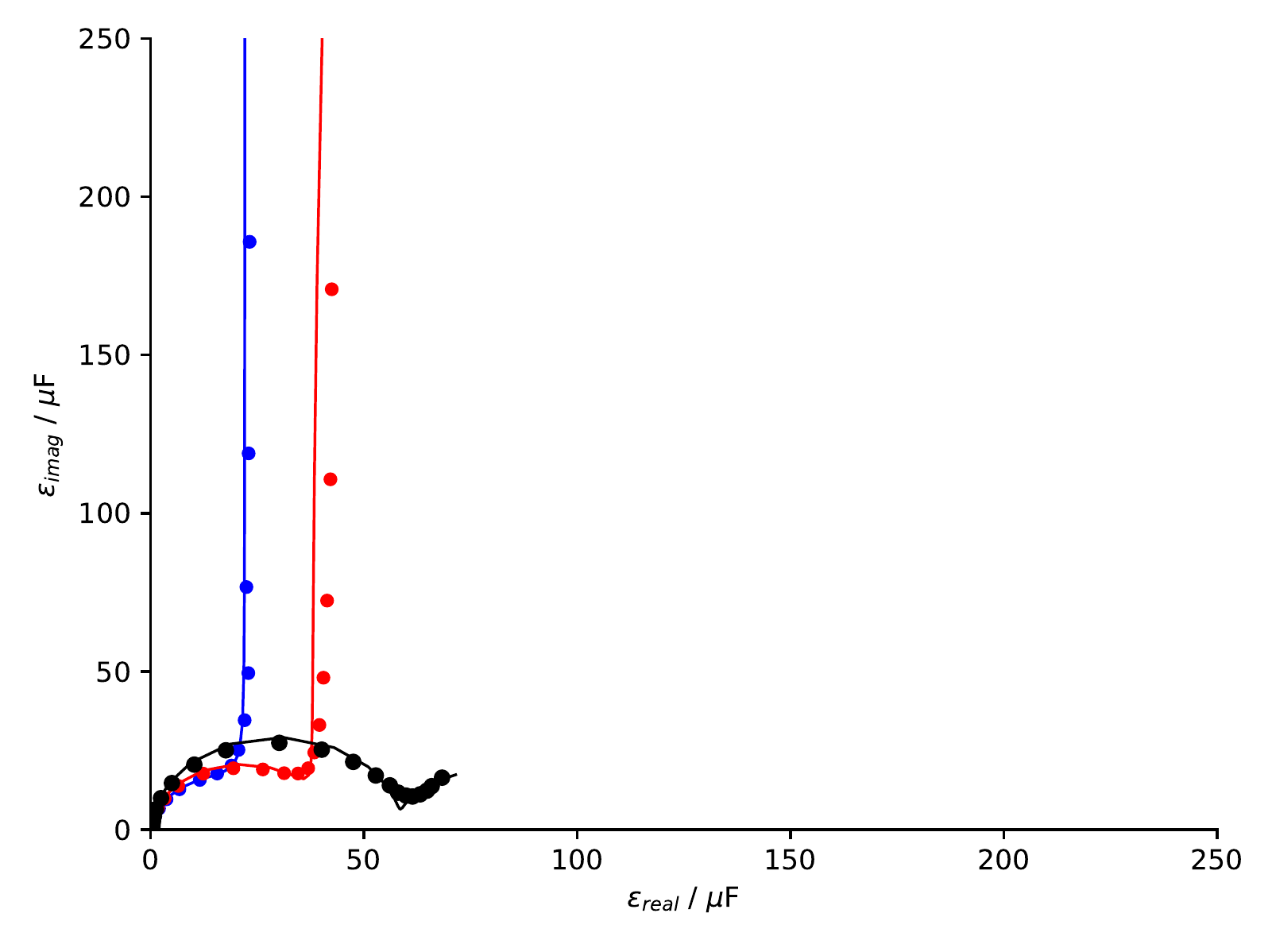}\\
\vspace{0.2cm}
{(c)}\\
  \caption{Upper panel: Complex permittivity plots of Ag(hkl) electrodes in 0.01 M H$_{3}$PO$_{4}$. Left: Spectra of Ag(111) electrode obtained at -0.20 V (black), -0.40 V (red), and -0.55 V (blue). Right: Spectra of Ag(100) electrode obtained at -0.20 V (black), -0.40 V (red), and -0.60 V (blue).
Middle panel: Complex permittivity plots of Ag(hkl) electrodes in 0.01 M KH$_{2}$PO$_{4}$obtained at -0.60 V (black), -0.70 V (red) and -0.85 V (blue). 
Lower panel: Complex permittivity plots of Ag(hkl) electrodes in 0.01 M K$_{3}$PO$_{4}$. Left: Spectra of Ag(111) electrode obtained at -0.80 V (black), -1.20 V (red) and -1.25 V (blue).Right: Spectra of Ag(100) electrode obtained at -0.60 V (black), -1.25 V (red) and -1.35 V (blue). 
Solid lines are the fitted curves. Upper panel: The $EEC$ in Fig. 1 (c) was used for all spectra.
Middle panel: The fit corresponding to $EEC$ (c) is shown for both surfaces at -0.60 V. $EEC$ in Fig. 1 (e) is shown for Ag(100) at -0.70 V and for Ag(111) at -0.85 V. $EEC$ (c) is shown for Ag(111) at -0.70 V. Lower panel: $EEC$ (d) is shown for Ag(100) at -1.30 V (blue). The fit using the model of $EEC$ in Fig. 1 (e) and (f) are shown for Ag(111) at -0.80 V (black) and -1.20 V (red), respectively.}
\end{figure}

For both electrodes a semicircle is observed at high frequencies, and
at potentials more negative than -0.55 V the response deviates from a
purely capacitive behaviour, being more evidenced for (100) surface.
The $EEC$ of Fig. 1 (b) gave inaccurate fitting of the data in the
whole potential range, corroborating previous results (see $SI$)
that diffusional processes were not the rate determining step \cite{salim2017}.
In the case of Ag(100) surface, the spectrum of -0.6 V deviates clearly
from a pure capacitance at low frequencies, demonstrating the coexistence of two simultaneous processes.
Therefore, from -0.60 V to -0.20 V
as the charge transfer is not taking place,
the $EEC$s in Fig. 1 (c) or (e) were employed.
A clear trend is observed for both surfaces, indicating that the potential of zero charge
($pzc$) is shifted and may be located in the potential region
where the $her$ occurs \cite{valette198237, valette1984179}.

In the case of Ag(hkl) in 0.01 M KH$_{2}$PO$_{4}$ (Fig. 2b),
from -0.60 V to -0.85 V a different behavior can be clearly seen
for the two electrodes. However,
all spectra were well fitted by
the $EEC$ of Fig. 1 (c),
as shown in the following section.
Fig. 2c shows complex permittivity plots of Ag(hkl) electrodes in 0.01 M K$_{3}$PO$_{4}$.
In the potential range between -1.0 V and -1.40 V the desorption of phosphate species
takes place, and differentiated responses of the (111) and (100) electrode surfaces
are obtained (see the $SI$ and the next section).
At -1.25 V, the fit corresponds to $EEC$ in Fig. 1 (e) for (100) surface.
$EEC$ (d) is shown for Ag(100) at -1.30 V (blue). The fit using the model
of $EEC$ in Fig. 1 (e) and (f) are shown for Ag(111) at -0.80 V (black),
and -1.20 V (red), respectively.
This fact demonstrates that the technique is sensitive to
the surface structure of the interface.

The dielectric complex permittivity representation of the data clearly 
demonstrates that for potentials more negative than the $her$ onset
the co-occurrence of processes associated to phosphate species takes place,
as the semicircle can be seen at all potentials.
However, for selected potentials, the difference
in the values of the associated time constants of
these two processes is more evident from the impedance spectra (insets of Fig. 2 left).

Comparisons and details of results using the $EEC$s of Fig. 1 are
provided in Sections 3.2 and 3.5.

\subsection{Cyclic voltammetry and R$_{}$($\phi$) plots}
Fig. 3 Upper panel shows votammogram profiles
of Ag(hkl) in 0.01 M H$_{3}$PO$_{4}$$_{(aq)}$,
starting the scanning at -0.20 V.
It shows that the hydrogen evolution
overpotential
is about 100 mV more
negative on Ag(111) as compared to Ag(100),
where a quasi-reversible
process between -0.70 V and -0.55 V
can be resolved.

\begin{figure}
\centering
  \includegraphics[height=4.5cm]{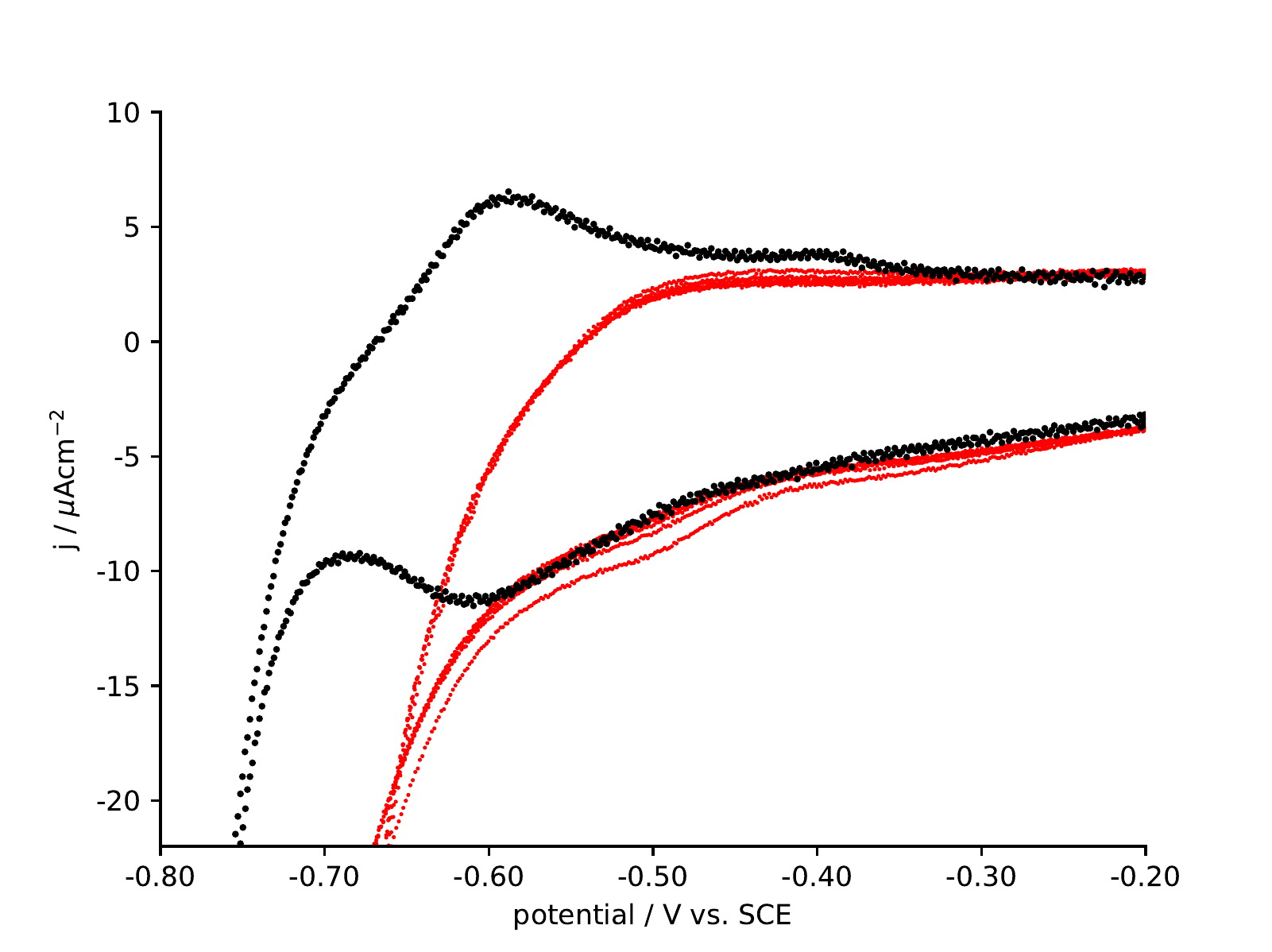}\\
\vspace{0.2cm}
\vspace{0.2cm}
  \includegraphics[height=4.5cm]{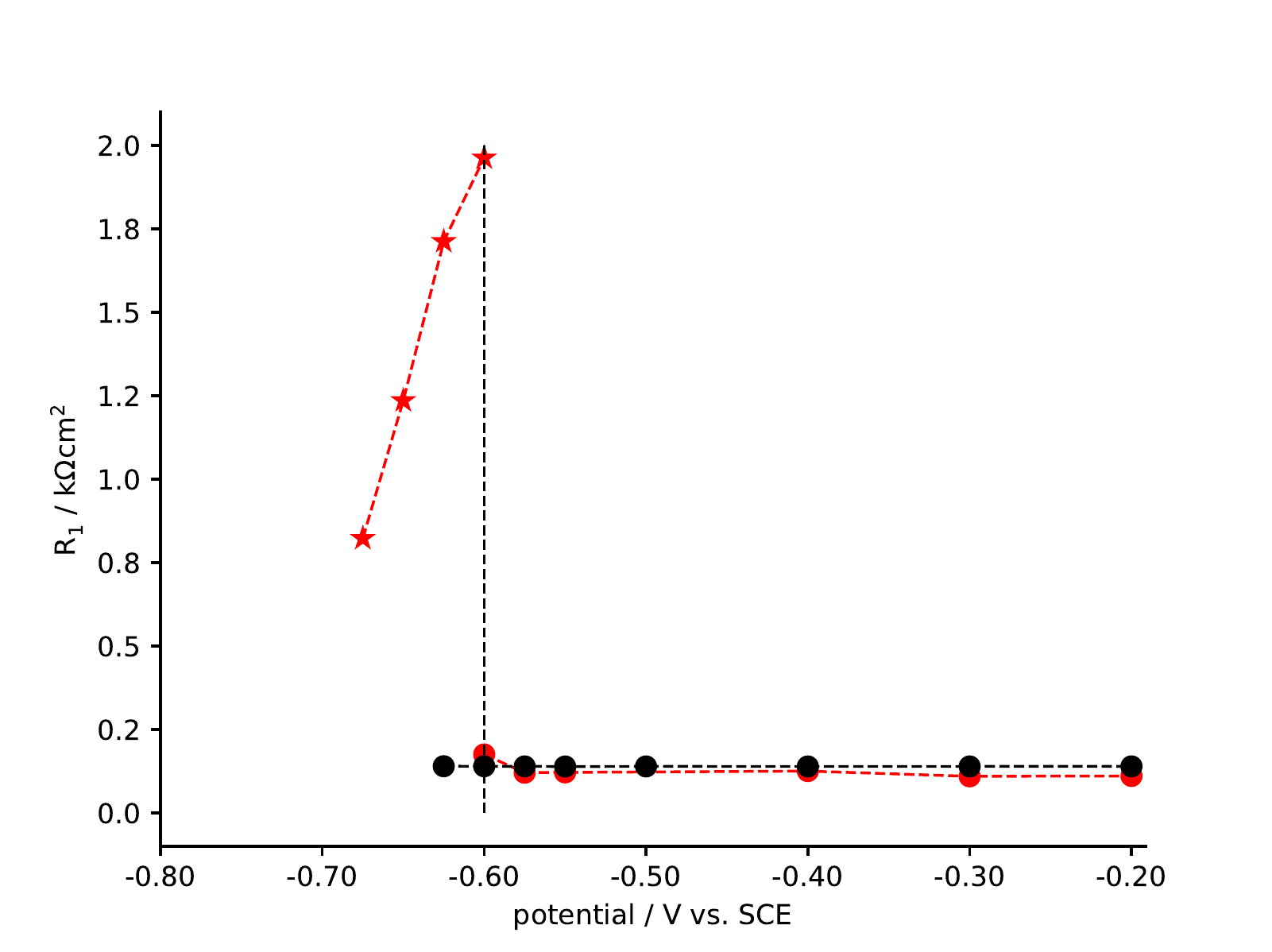}\\
\vspace{0.2cm}
  \caption{Upper panel: CVs plots of Ag(111) (black) and Ag(100) (red) electrodes in 0.01 M H$_{3}$PO$_{4}$. 
Scan rate 0.05 Vs$^{-1}$. 
Lower panel: R$_{1}$ (black and red solid circles) and R$_{ct}$ (red stars) as a function of potential. 
R$_{1}$ and R$_{ct}$ were obtained from the fitting parameters for the impedance plots. The vertical dashed line indicates the potential region where R$_{ct}$ values were obtained using $EEC$ in Fig. 1 (d) (potentials more negative than -0.60 V, left side) and R$_{1}$ using $EEC$ in Fig. 1 (c) (potentials more positive than -0.60 V, right side).}
\end{figure} 

The study of the
onset potential of $her$
as a function of pH on Ag(hkl)
was first reported by
Diesing et al. by means of
surface resistance measurements \cite{diesing1997}.
They have found
that the two faces
have the same
onset potential of $her$
in 0.1 M HClO$_{4}$, but
differ in neutral solution.
They have concluded that in 0.1 M KClO$_{4}$
$her$ is based on water dissociation.
As the only difference with our experimental
conditions is the electrolyte used,
it is expected that the differences observed in the onset potential of $her$
with respect to our results are directly associated
with the presence of the adlayer of phosphate species.

Fig. 3 Lower panel shows the R$_{1}^{}$ as a function of potential.
From -0.55 V towards positive potentials
the R$_{1}^{}$ values
were calculated from
the $EIS$ data fitting
using the $EEC$ in Fig. 1 (c) for both surfaces. The R$_{1}^{}$
remains almost constant
with low values.
In the case of Ag(100) surface, the vertical dashed line in the figure separates two
characteristic potential regions, indicating
that the $EEC$ in Fig. 1 (d) was used in the low overpotential for $her$, where it has physical significance.
When comparisons were made with the other $EEC$s, this $EEC$ was fitted
to the data with the lowest values of $\chi^{2}$ and relative percentage errors (see Table S6).
The R$_{ct}$ is larger than the values found for R$_{1}$, which is a
reasonable result due to both low electrolyte concentration and low overpotentials.

On silver electrodes, $her$ takes place
through a two-step mechanism, involving Volmer and Heyrovsky
as elementary steps. In acid solutions at high overpotentials,
the reaction proceeds via Volmer-Heyrovsky mechanism, and the first
step is the hydrogen atom adsorption \cite{ruderman2013}.
Thus, as observed from the voltammograms, the hydrogen atom should
be adsorbed on the surface at least from -0.6 V for the
(100) and -0.7 V for the (111).
This is an important finding that follows the same trend
as the results reported by Kolovos-Velianitis, \cite{kolovos2004}
and opposite to the results found by Kuo et al.,
where DFT calculations and experiments with nanoparticles
using 0.5 M H$_{2}$SO$_{4}$ and 1 M ethanol as electrolyte
have shown that the (111) facet
has higher activity for $her$ than (100) facet \cite{kuo2019}.

Fig. 4 Upper panel shows the cyclic voltammograms of
Ag(hkl) electrodes in 0.01 M KH$_{2}^{}$PO$_{4}^{}$,
starting the scanning at -0.20 V.
For both systems the behavior observed
in the hydrogen evolution region shows
a significant hysteresis
when scanning the potential between
negative and positive directions,
and the onset of $her$
starts at around -0.90 V for both electrodes.
For the Ag(111) surface there is a potential region
where the current is constant
before $her$.
It should be note that at this pH (4.79) the current
density increasement is not as sharp as at 1.60.

Fig. 4 Lower panel shows the R$_{1}$ values obtained using the $EEC$ in Fig. 1 (c)
(black and red circles) from -0.80 V to -0.60 V for both electrodes, and the $EEC$
(e) between -0.55 V and -0.20 V for Ag(100) (R$_{1}$ and R$_{2}$, red stars).
The R$_{1}$ values obtained using $EEC$ (e) follow the same trend as for $EEC$ (c),
and compared with R$_{2}$ values, these are almost two orders of magnitude larger.
The fact that $EEC$ (c) gave the best fitting results (see $SI$) in the potential region where
the current peaks in the $j-V$ profiles are observed, indicates that the $EIS$
response can be associated with only one predominant process, which is highly
reversible (Fig. 4 upper panel).
Further discussion is provided in sections 3.3 and 3.5.

\begin{figure}
\centering
  \includegraphics[height=4.5cm]{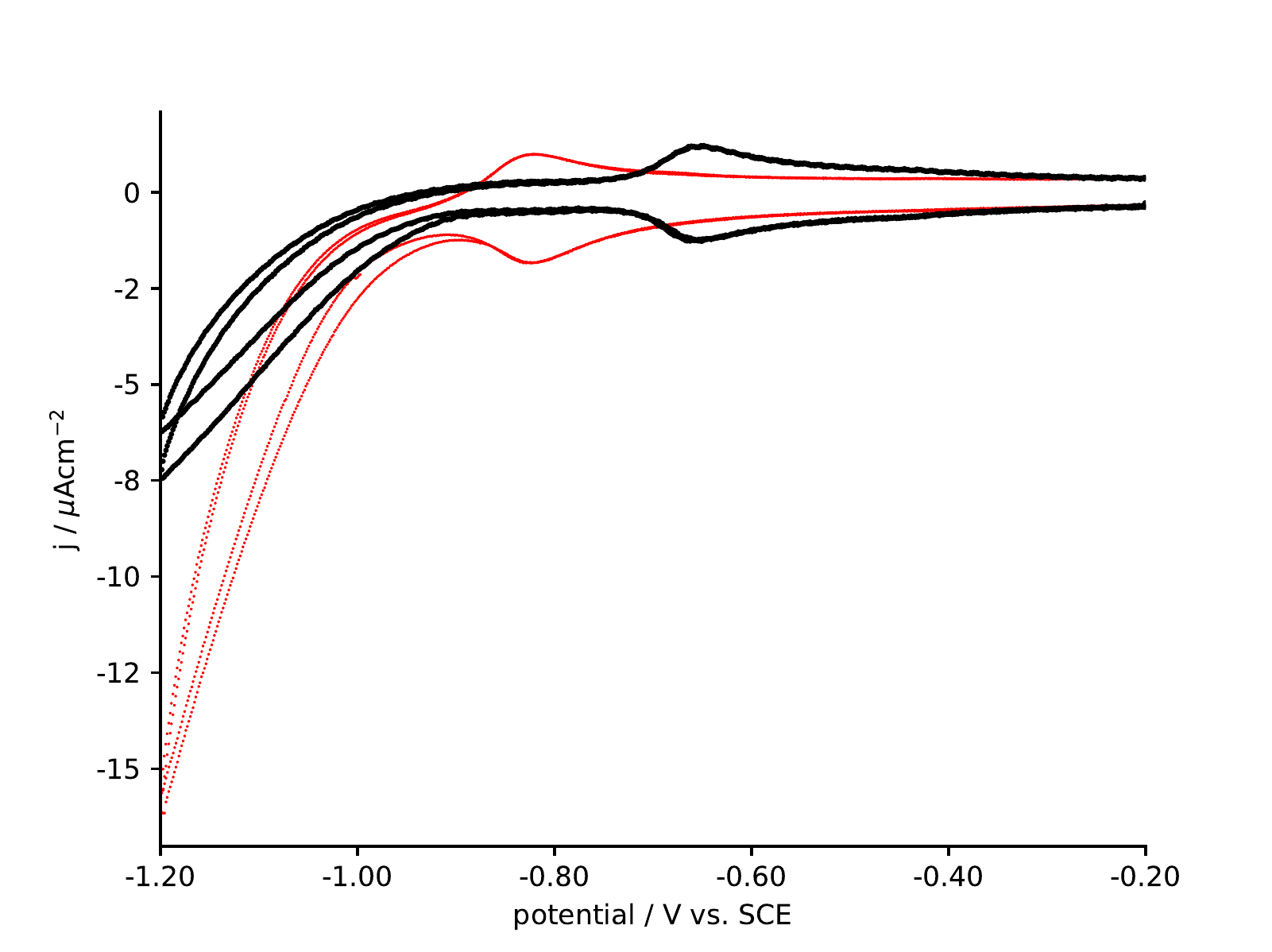}\\
  \includegraphics[height=4.5cm]{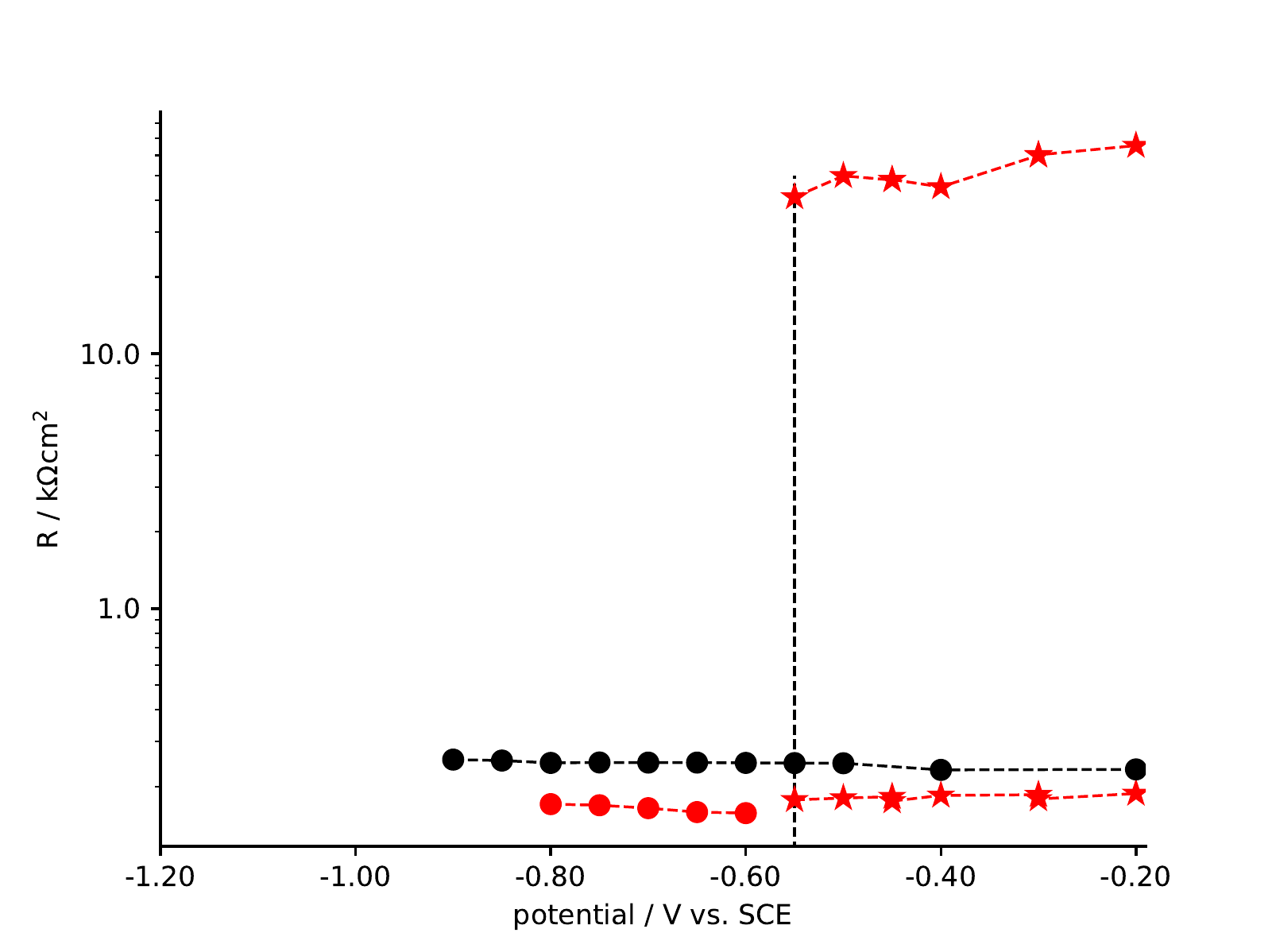}\\
\vspace{0.2cm}
  \caption{Upper panel: CVs plots of Ag(111) (black) and Ag(100) (red) electrodes in 0.01 M KH$_{2}$PO$_{4}$. 
Scan rate 0.05 Vs$^{-1}$. 
Lower panel: R$_{1}$ as a function of potential obtained using $EEC$ in Fig. 1 (c) (black and red circles). The vertical dashed line at -0.55 V indicates that the $EEC$ in Fig. 1 (e) gave the best fitting from -0.55 V to -0.20 V for the Ag(100) electrode (R$_{1}$ and R$_{2}$, red stars). (See text and $SI$ for explanation).}
\end{figure}

Fig. 5 Upper panel shows the cyclic voltammograms of
Ag(hkl) electrodes in 0.01 M K$_{3}$PO$_{4}$,
starting the scanning at -0.20 V.
The peaks observed for phosphate species at potentials more positive than -1.25 V
indicate that different adsorbates are present on Ag(hkl) with different bond energies.
In the potential region where $her$ takes place,
hysteresis is not observed in either of the two orientations.
At pH=12, the onset potential of $her$ occurs about -1.25 V for
(100) surface, where the current increase is sharp and well defined,
contrasting with (111) face.
The inset shows where the adsorption (-1.20 V) and desorption (-1.25 V) peaks
are well defined for the Ag(111) in
0.1 M K$_{3}$PO$_{4}$. Under these conditions
the phosphate species desorption process can be clearly detected.

The resistance parameters obtained using $EEC$s (c), (d), (e) and (f) are shown in Fig. 5 Lower panel
as a function of potential.

The plot on the left shows the values of R$_{1}$ and R$_{ct}$ for both electrodes.
The vertical dashed line shows that from -1.40 V to -1.20 V, the R$_{ct}$ was obtained by $EEC$ (f) (stars), and from -1.20 V to -0.20 V, R$_{1}$ was obtained by $EEC$ (e) (circles).
Therefore, from these results in the two potential regions, the physical model represented by the $EEC$ (e) can be associated with the coexistence of two processes for potentials positive to -1.20 V. On the other hand, at potentials less positive than -1.20 V, the $EEC$s (d) and (f) were used that consider the co-occurrence of the charge transfer. The $EEC$ shown in Fig. 1 (d) is a simplification of that in Fig. 1 (f) when the participation of one of the two adsorption processes is negligible. However, in this case the best fitting of the results was obtained using the $EEC$ shown in Fig. 1 (f) (see Tables S8 and S9). Morin et al. \cite{morin1996} have studied the kinetics of $upd$ of H on Pt(hkl) in 0.5 M H$_{2}$SO$_{4}$ by $EIS$, and the same $EEC$ as in Fig. 1 (e) was associated with a kinetic model that considers the presence of anion co-adsoption.

In order to compare how resistances are modified in the presence
and absence of phosphate species, the results obtained for the Ag(111)
in 0.01 M KOH (blue circles) are also shown (right side plot).
It is observed that lower values of R$_{2}$ are obtained in
presence of phosphate species in the whole potential range, i.e.,
the rate of OH$^{-}$ adsorption processes increases.

\begin{figure}
\centering
  \includegraphics[height=4.5cm]{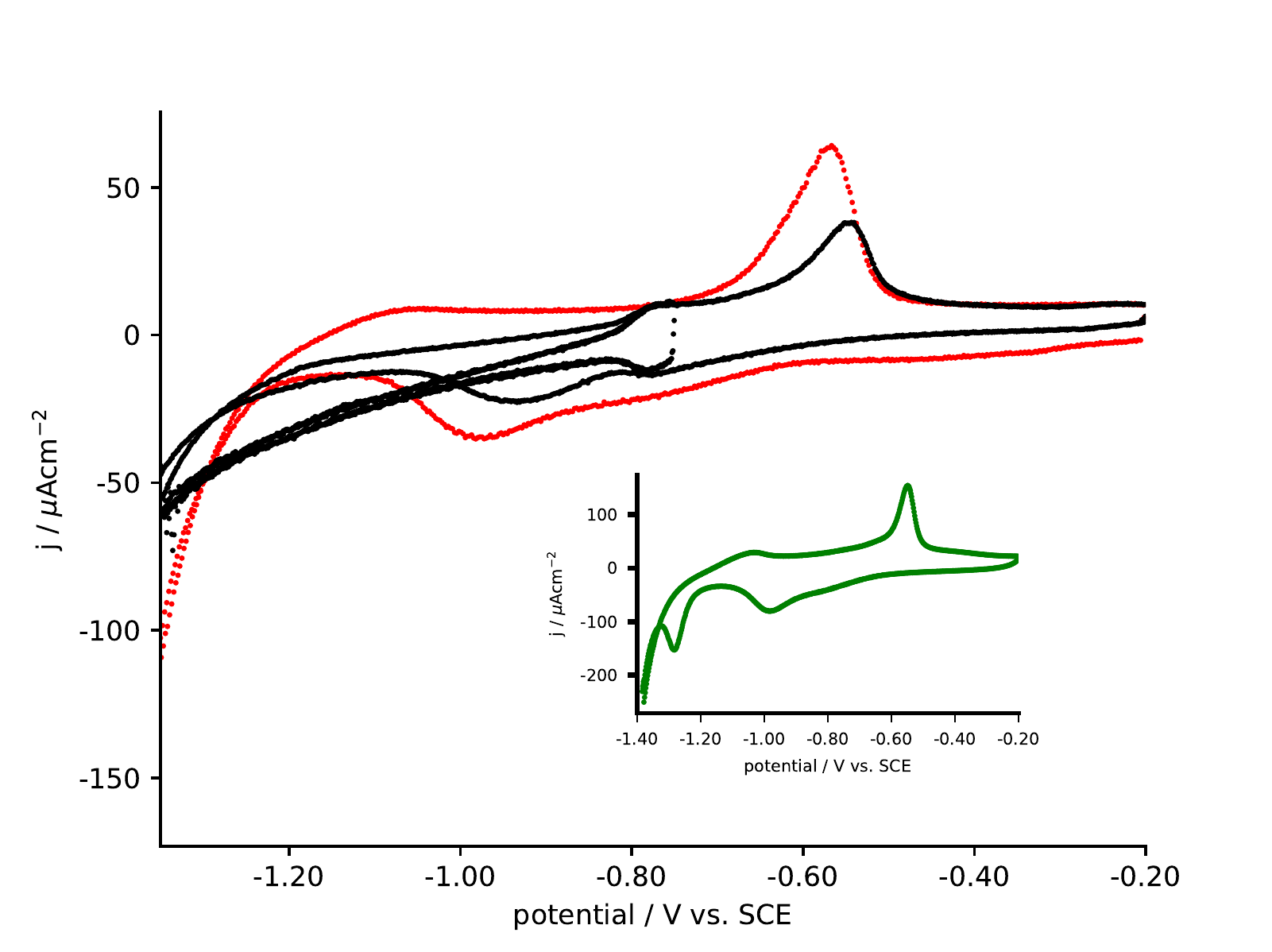}\\
  \includegraphics[height=4.5cm]{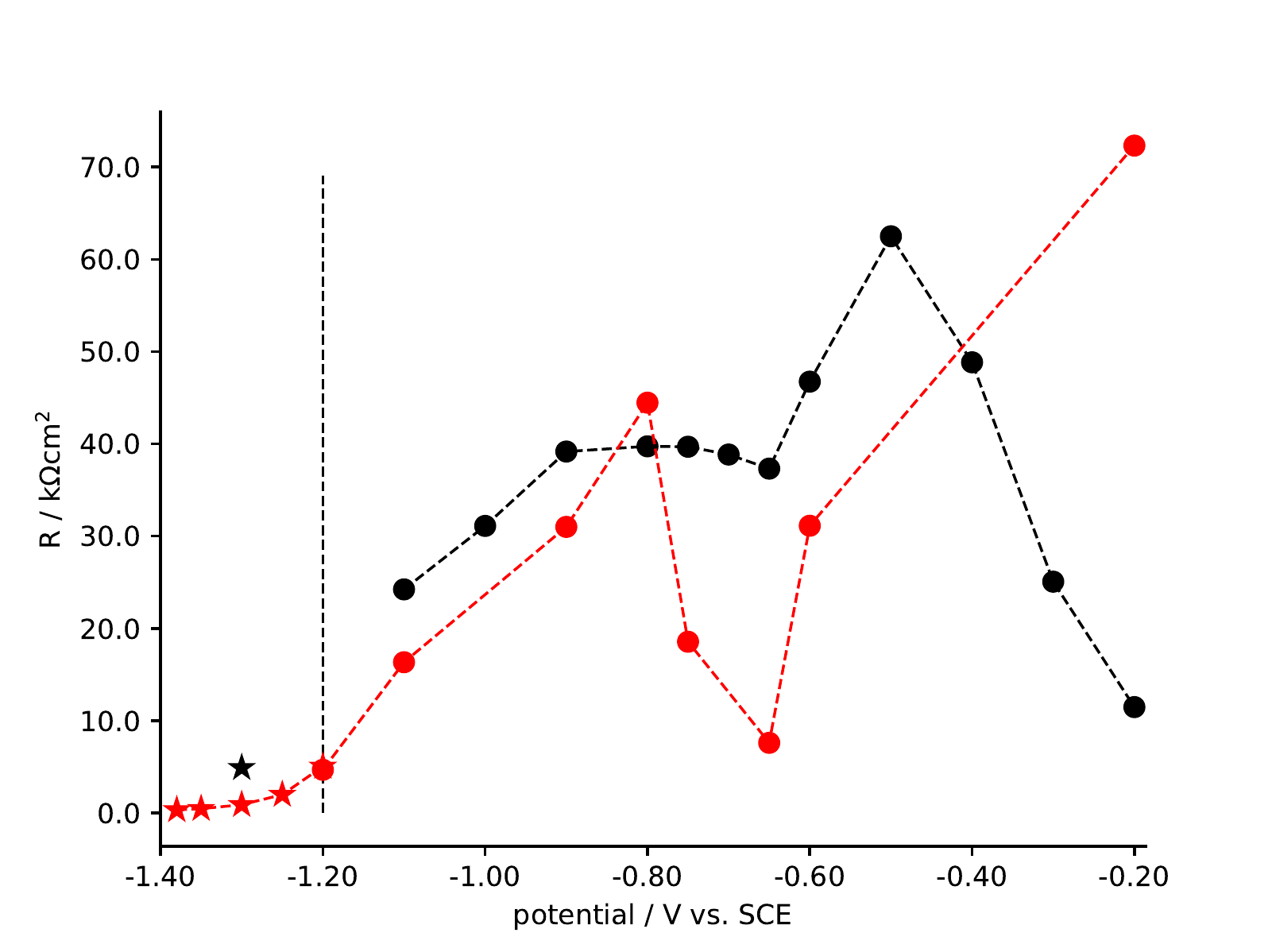}
  \includegraphics[height=4.5cm]{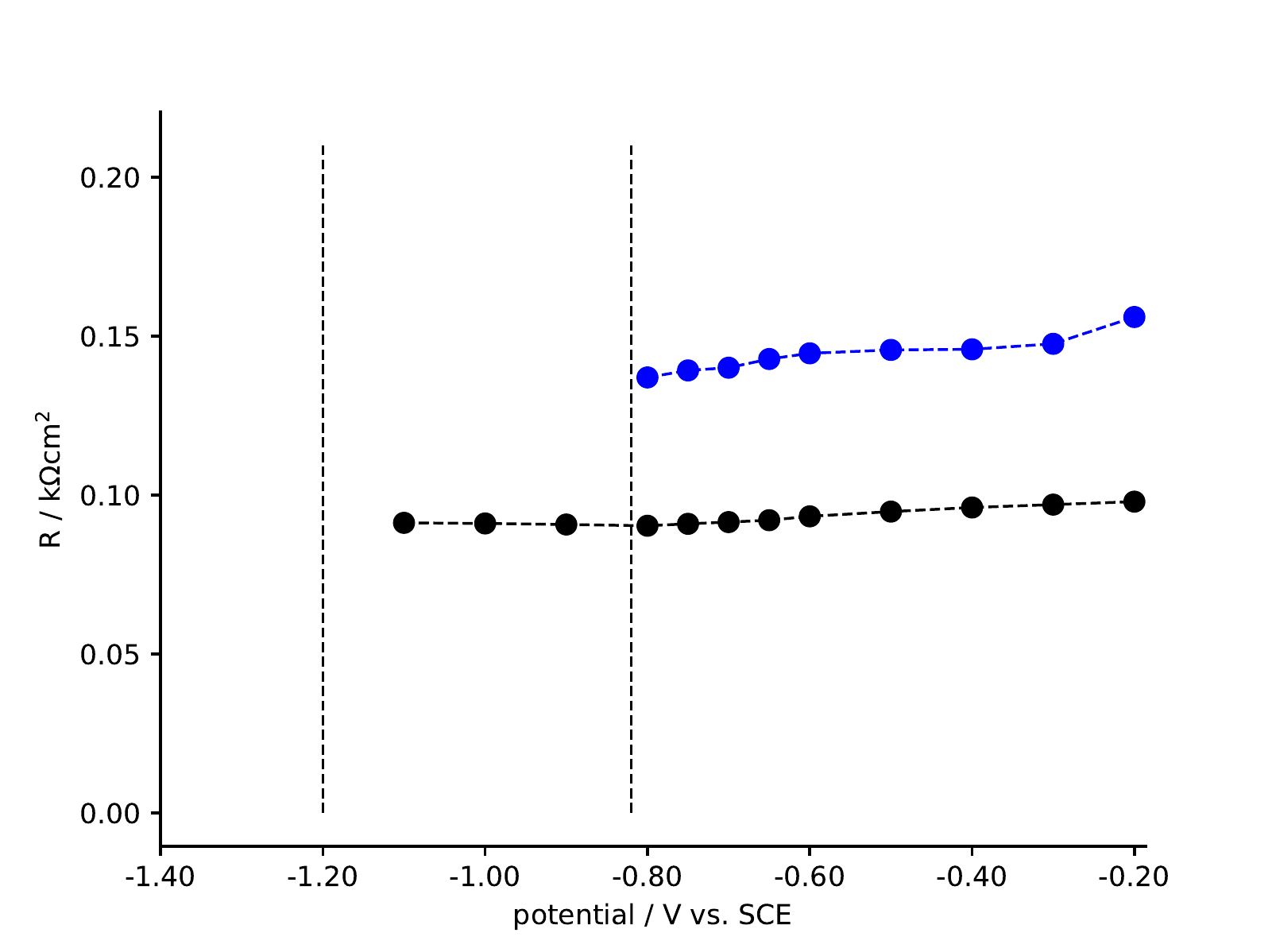}\\
\vspace{0.2cm}
  \caption{Upper panel: $CVs$ plots of Ag(111) (black) and Ag(100) (red) electrodes in 0.01 M K$_{3}$PO$_{4}$. 
A voltammogram of Ag(111) electrode starting at -1.40 V up to -0.75 V (black) is also included. Scan rate 0.05 Vs$^{-1}$. 
Inset: $CV$ plot of Ag(111) electrode in 0.1 M K$_{3}$PO$_{4}$. Lower panel: Left: R$_{1}$ (circles) and R$_{ct}$ (stars) parameters obtained by $EIS$ data fitting to the $EEC$s in Fig. 1 (e) at potentials positive to -1.20 V and (f) at potentials negative to -1.20 V, as shown by the vertical dassed line. Right: R$_{2}$ parameters (black circles) obtained by fitting to the $EEC$ in Fig. 1 (e) for Ag(111) electrode in 0.01 M K$_{3}$PO$_{4}$, and R$_{1}$ parameters (blue circles) obtained by fitting to the $EEC$ in Fig. 1 (c) for Ag(111) electrode in 0.01 M KOH. The vertical dashed line at -0.82 V shows correspondence with the potential of the current onset in the voltammogram associated with OH$^{-}$ species.}
\end{figure}

A voltammogram of Ag(111) electrode starting at -1.40 V up to -0.75 V
is also included to identify the OH$_{(ads)}$ transition to the ordered layer (vertical dashed line at -0.82 V),
which for this face is well differentiated from phosphate species processes \cite{salim2017, horswell2004}.
Therefore, the two species that can be resolved by $EIS$ and $CV$ results were associated with PO$_{4}$$_{(ads)}$ and OH$_{(ads)}$, corresponding to Eq. (3) as discussed in section 4.

\subsection{C($\phi$) plots for Ag(hkl)/H$_{2}$PO$_{4}$$_{(ads)}$}
Fig. 6 shows C$_{1}$($\phi$) vs. potential curves
of Ag(111) (black) and Ag(100) (red) electrodes
in 0.01 M KH$_{2}$PO$_{4}$.
For comparison the inset shows the differential capacitance curves obtained in 0.01 M KClO$_{4}$
where the $pzc$ is -0.850 V and -0.735 V for (100) and (111) silver surfaces, respectively.

C$_{1}$($\phi$) were obtained from $EEC$ in Fig. 1 (c). In the case of (100) surface, from -0.50 V to -0.3 V the $\chi^{2}$ and \%errors were high (see Table S5). However, very similar values were obtained using $EEC$ (e).

The highest values of C$_{1}$($\phi$) are
found at -0.60 V and -0.77 V
for Ag(111) and Ag(100), respectively. However,
it is observed that the shape of C$_{1}$($\phi$) curves
in 0.01 M KH$_{2}$PO$_{4}$
is not face-specific \cite{hamelin1985modern}.
The only difference observed is the x-axis shift,
as expected according to the adsorbate-surface bond energy.
Table 1 shows the capacity values that correspond to the peak
in the range of electrode potentials giving positive charge density.
It can be seen that 135 $\mu$Fcm$^{-2}$ was obtained
for Ag(111)/phosphate species, whereas less than
130 $\mu$Fcm$^{-2}$ was obtained for Ag(100).
Other adsorbates are also shown for comparison \cite{doubova1997785}.
For example, Beltramo et al. \cite{beltramo2003127} have reported
near 90 $\mu$Fcm$^{-2}$ and 150 $\mu$Fcm$^{-2}$ in 0.1 mM for Ag(111)/chloride species
and for Ag(111)/bromide species, respectively, whereas at 10 mM,
120 $\mu$Fcm$^{-2}$ and 180 $\mu$Fcm$^{-2}$ values were found, respectively (see Table 1).
Therefore this denotes a strong dependence on
the nature and concentration of the adsorbate,
which determines the redistribution of the charge
density at the interface \cite{huang2019, bazant2019A1},
consequently affecting the co-adsorption of other species.

\begin{table}[htb]
\centering
\begin{tabular}{c c c c}
\qquad \qquad \qquad &   &   &  \\
\hline
Adsorbate \qquad \qquad & Ag(111)  & Ag(100) &  Ref.\\
\hline
H$_{2}$PO$_{4}^{-}$     & 135  & 127 & this work\\ 
HPO$_{4}^{2-}$, PO$_{4}^{3-}$      & 160 & 110 & this work\\
OH$^{-}$      & 8  & -- & this work\\
CH$_{3}$COO$^{-}$     & 8  & -- & (39)\\
SO$_{4}^{2-}$      & 15  & -- & (39)\\
F$^{-}$     & 5, 80-90 (30 - 40 mM)  & -- & (48, 34)\\
Cl$^{-}$ (0.1 mM)    & 36  & -- & (24)\\
Br$^{-}$ (0.1 mM)     & 60  & -- & (24)\\
\hline
\end{tabular}
\caption{Values of capacity maximum in ranges of potential of
positively charged surface.
The C$_{1}$ are in $\mu$F cm$^{-2}$. Concentration of anions in the bulk solution when not specified are 10 mM.}
\end{table}

The characteristic shape observed for the plots
of C$_{1}$($\phi$) as a function of potential
has been consistently studied by several authors
\cite{valette198237, valette1989191, hamelin1985modern, valette1981285, valette1987189water, hamelin198915}.
They have discussed the shape and potential shift of these curves
and have found for gold electrodes in halide solutions
that the $pzc$ shifts towards more negative potential values
as the anion size becomes bigger, together with higher values of the capacitance of
the adsorption peaks and decrease of their widths.

\begin{figure}
\centering
  \includegraphics[height=4.5cm]{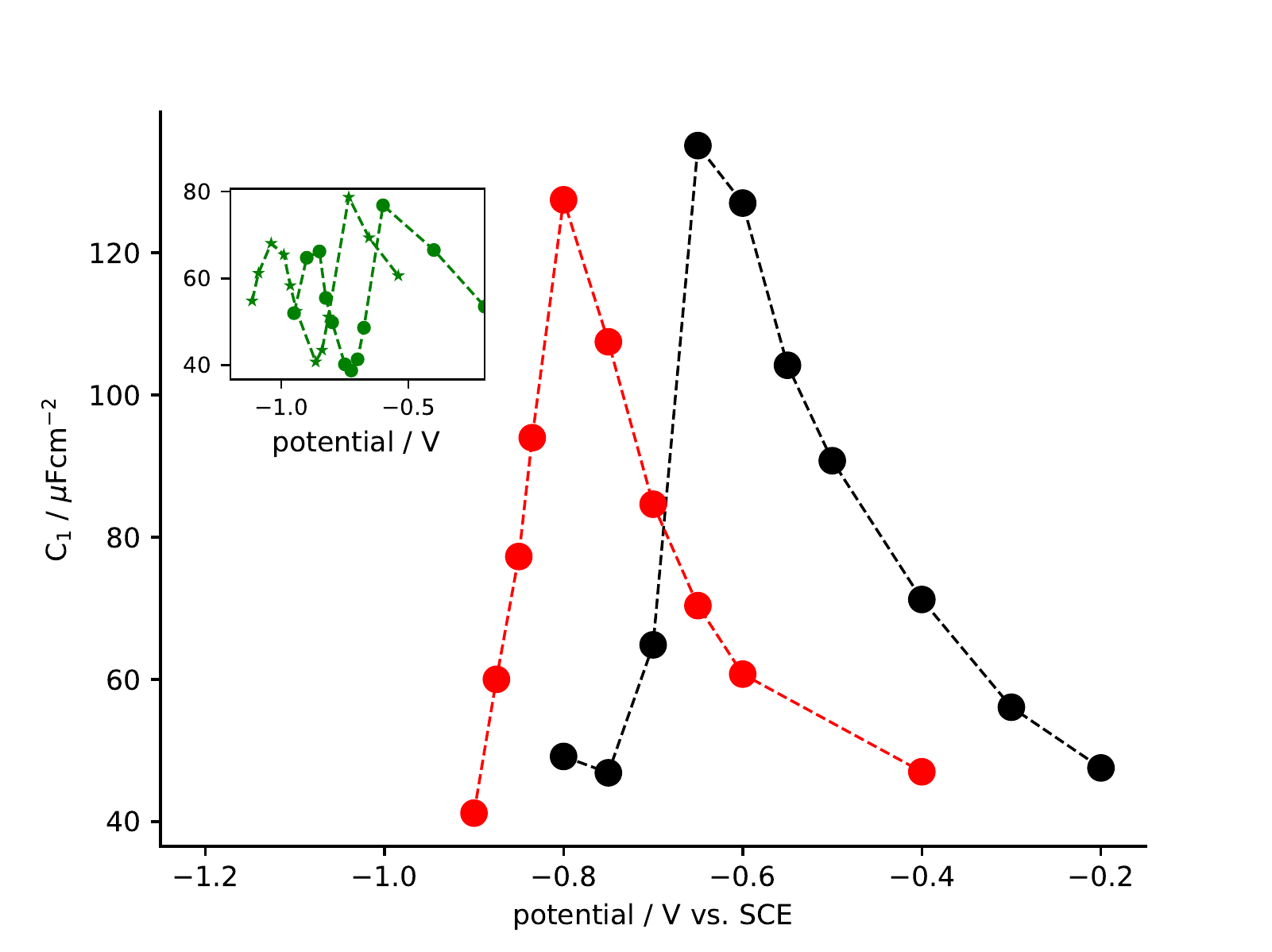}\\
  \caption{C$_{1}$($\phi$) vs. potential plots of Ag(111) (black) and Ag(100) (red) electrodes in 0.01 M KH$_{2}$PO$_{4}$. 
 Inset: Similar plots of both electrodes obtained in 0.01 M K$_{}$ClO$_{4}$ (green) are also included for comparison. Ag(111) (solid circles) and Ag(100) (stars). Dashed lines are for visual reference.}
\end{figure}

\subsection{C($\phi$) plots for Ag(hkl)/PO$_{4}$$_{(ads)}$}
Fig. 7 Upper panel shows C$_{1}$($\phi$) vs. potential curves
of Ag(111) (black) and Ag(100) (red) electrodes
in 0.01 M K$_{3}$PO$_{4}$, which are markedly different for the two
electrodes in the whole potential range.
C$_{1}$($\phi$) values were obtained using the $EEC$s in Fig. 1 (c) and (e).
In this case the $[$HPO$_{4}^{2-}$/PO$_{4}^{3-}$$]$ $\approx$ 10
and the pH = 12.

A wide minimum between -1.0 and -0.8 V
is observed for Ag(100) surface.

\begin{figure}
\centering
  \includegraphics[height=4.5cm]{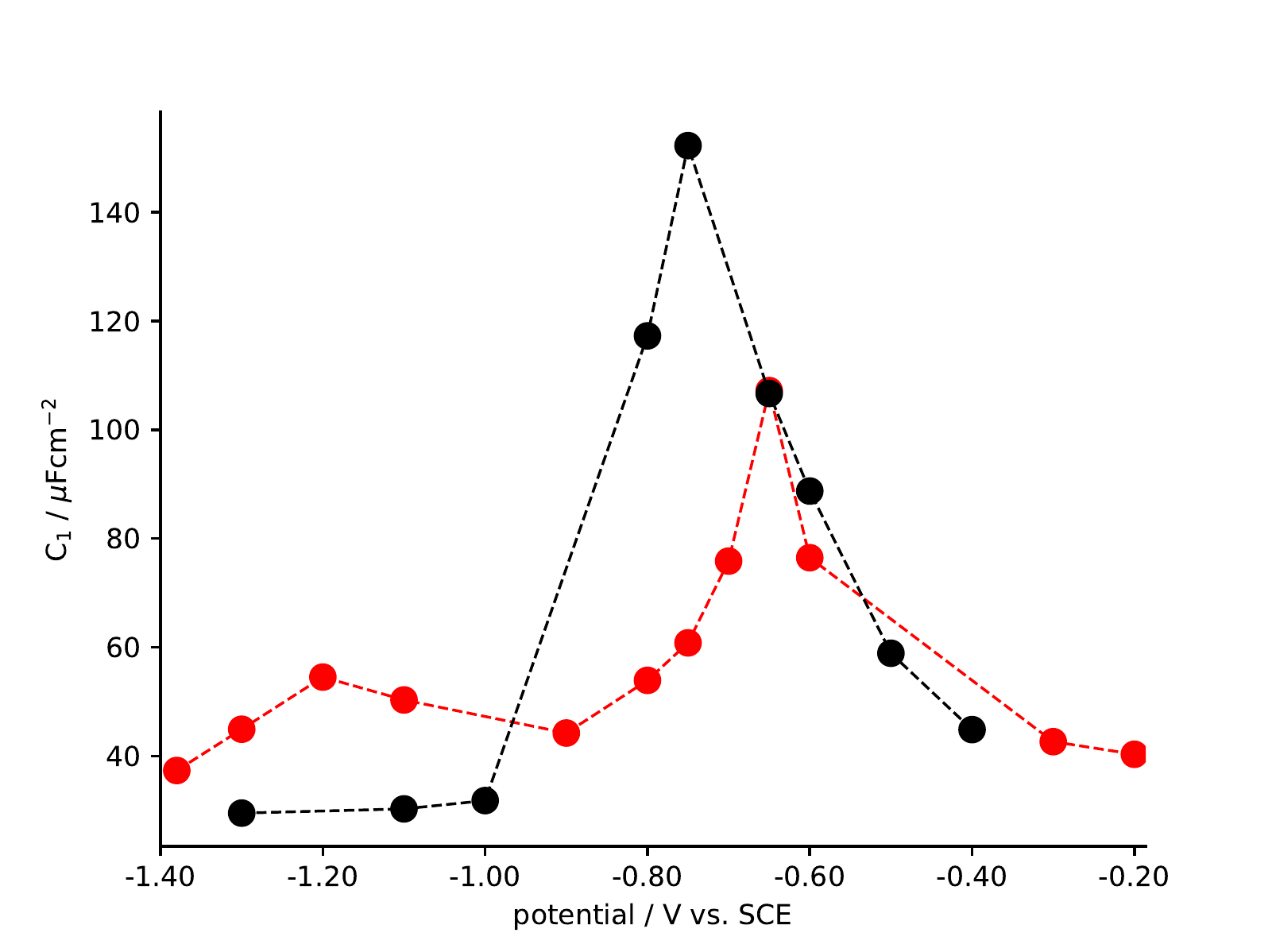}\\
  \includegraphics[height=4.5cm]{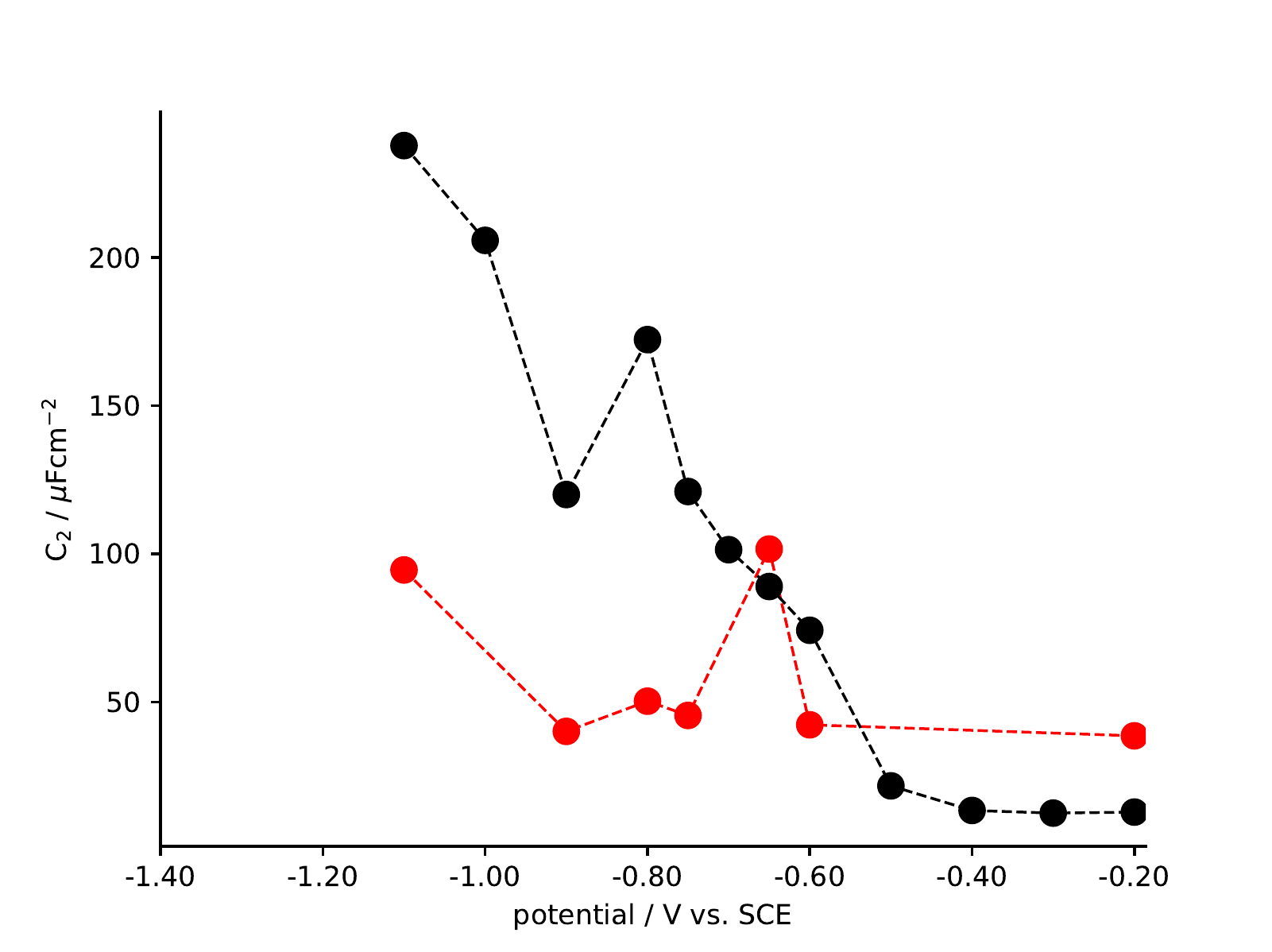}\\
   \caption{Capacitance$_{}$ vs. potential plots of Ag(111) (black) and Ag(100) (red) electrodes in 0.01 M K$_{3}$PO$_{4}$. 
Upper panel: C$_{1}$($\phi$) values obtained by the $EEC$ in Fig. 1 (c). Similar values of 
C$_{1}$($\phi$) were obtained using the $EEC$ in Fig. 1 (e). Lower panel: C$_{2}$($\phi$) 
values obtained by the $EEC$ in Fig. 1 (e) from -1.1 to -0.20 V. 
Dashed lines are for visual reference.}
\end{figure}

The peak shape of the C$_{1}$($\phi$) vs. potential curves
for (100) face is narrower than for the (111) face, following
a similar trend observed for sulfate ions on Ag(hkl) \cite{gao1994, hamelin198915}.
At potentials more negative than -1.25 V, the fitting of the $EEC$ in Fig. 1 (c)
for the analysis of $EIS$ spectra gave relative errors between 13\% and 20\% and
the $\chi^{2}$ $>$ 1. However, C$_{1}$($\phi$) values were almost equal to those
obtained with the $EEC$ in Fig. 1 (e), which gave better fitting results.
The C$_{2}$($\phi$) values (Fig. 7 Lower panel) at potentials more negative than
-1.20 V gave high relative errors (see Table S8), thus they are not shown on the graph.
Note that C$_{1}$($\phi$) values change with potential in the same way as the
current density in the $j-V$ profiles for both surfaces.
Therefore, when the $EEC$ of Fig. 1 (e) is applied, separation of two processes is possible.
The peaks of C$_{1}$($\phi$) for the (100) face (approximately at -1.20 V and -0.65 V)
correspond to minimum values of the R$_{1}$ observed in Fig. 5 Lower panel left, such
as predicted in Ref. \cite{jovic1995}. In contrast, for the Ag(111)/K$_{3}$PO$_{4}$ system,
no direct associations could be established between a minimum in R$_{1}$ and a
maximum in C$_{1}$$_{}$($\phi$).

\section{Discussion}

\subsection{Dynamical processes} 
Fig. 8 (upper panel) shows the
$\tau_{1}$ as a function of potential
of Ag(111) and Ag(100) electrodes in 0.01 M H$_{3}$PO$_{4}$, where the
negative limit of potential is -0.625 V and -0.600 V for (111) and (100), respectively.

\begin{figure}[h]
\centering
  \includegraphics[height=4.50cm]{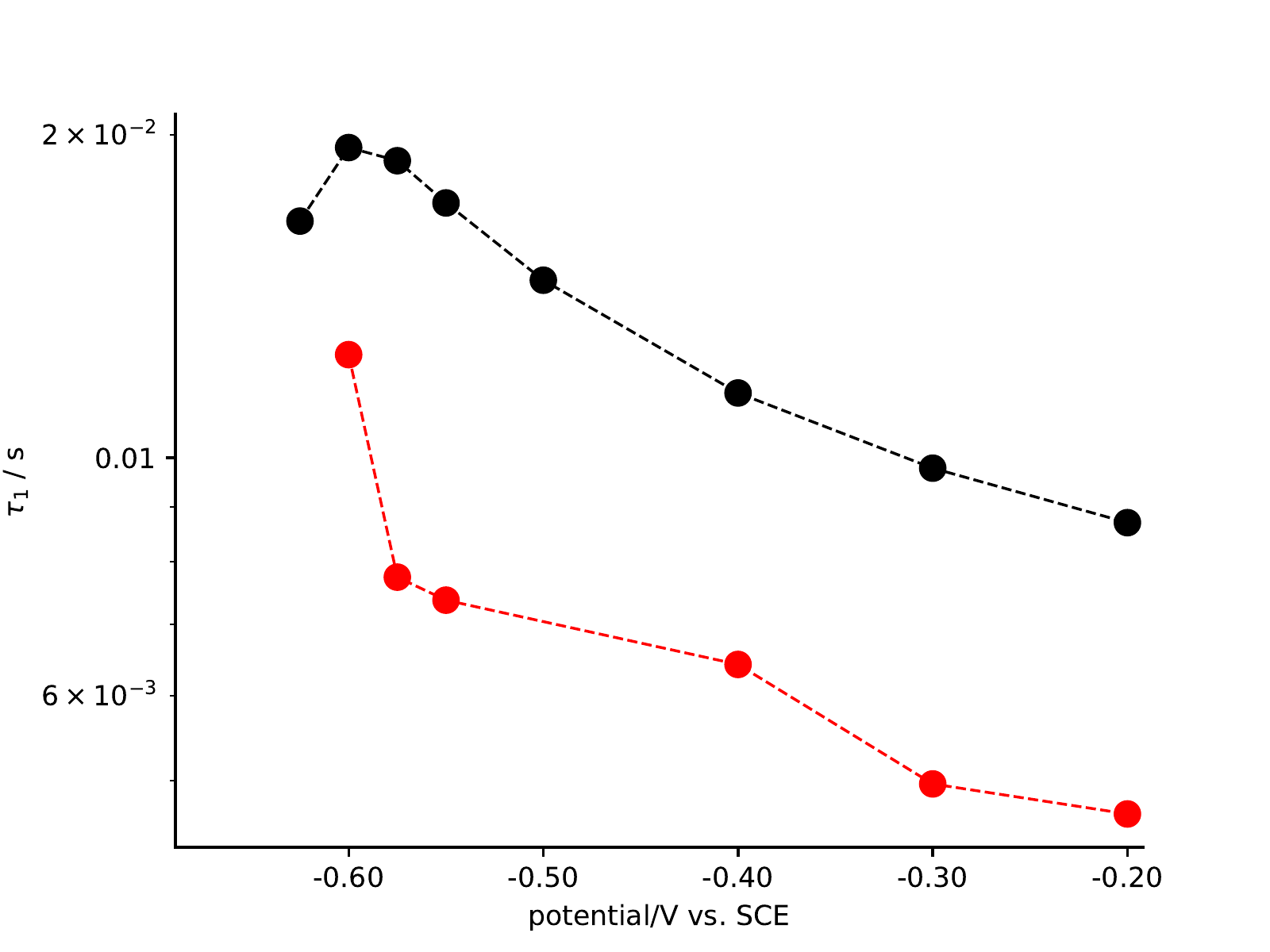}\\
  \includegraphics[height=4.50cm]{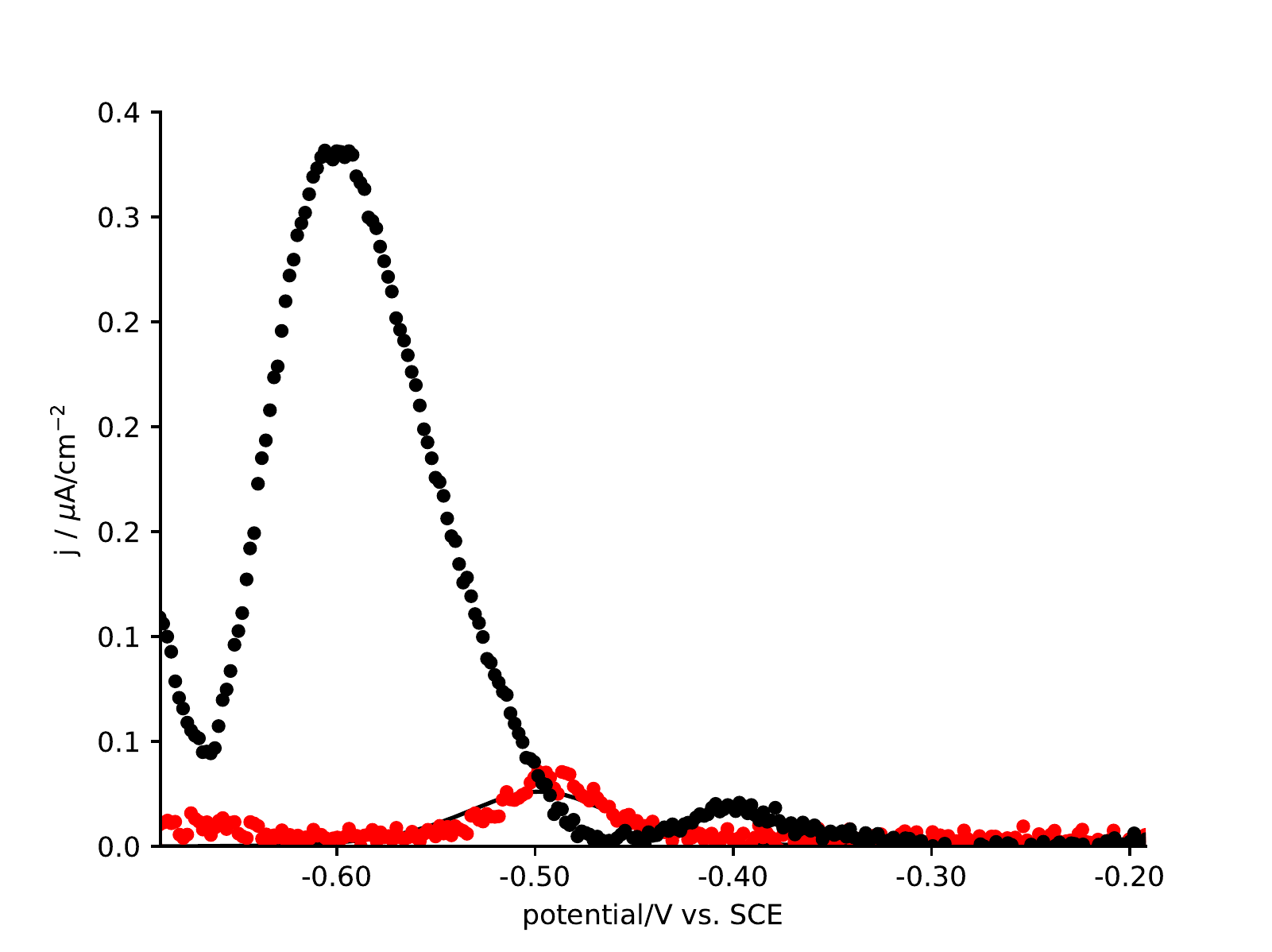}\\
  \caption{Upper panel: $\tau_{1}$ of Ag(111) (black) and Ag(100) (red) electrodes in 0.01 M H$_{3}$PO$_{4}$. 
The $\tau_{1}$ values were calculated by the product of R$_{1}$ and C$_{1}$ components of the $EEC$ in Fig. 1 (c). 
Lower panel: Anodic $j-V$ profiles. Scan rate 0.05 V/s. The graph shows only the anodic profiles 
for both electrodes (see section 1 in the $SI$).}
\end{figure}

Shorter $\tau$s were calculated for (100) face, whose values range between
3-10 ms. As noted in section 3.1, in the region of potential referred
to in this section, restructuration of the adsorbed layer is the most likely process.

Fig. 8 (lower panel) shows the anodic $j-V$ profiles of Ag(hkl) electrodes
in 0.01 M H$_{3}$PO$_{4}$, and Table S1 shows the adsorption charge of phosphate
species on the (111) face obtained from integration of anodic/cathodic peaks at -0.6 V (see $SI$).
Table 2 shows the coverage calculated considering the surface atom density of Ag(111)
orientation (1.38$\times$10$^{15}$ atoms/cm$^{2}$). The values seen correlate well with a low coverage.
On the other hand, we have found a value of 7.25 $\mu$Ccm$^{-2}$ calculated theoretically
by DFT for a p(4x4) ordered commensurate submonolayer of H$_{2}$PO$_{4}$$_{(ads)}$ adsorbed
on Ag(111) (coverage of 0.0625 ML) \cite{salim2017}.
We can conclude that a good agreement was obtained,
taking into account that the theoretical result
was performed under non-electrochemical
conditions, and that the experimental result
is subject to the error of numerical integration.

Fig. 9 (upper panel) shows the $\tau$ as a function of potential
of Ag(111) (black) and Ag(100) (red) electrodes in 0.01 M KH$_{2}$PO$_{4}$
between -0.80 V and -0.20 V.
At -0.90 V, $her$ onset is observed for both electrodes, consequently, for
this potential, the best fitting was given by $EEC$ (d) (see $SI$).

\begin{figure}
\centering
  \includegraphics[height=4.50cm]{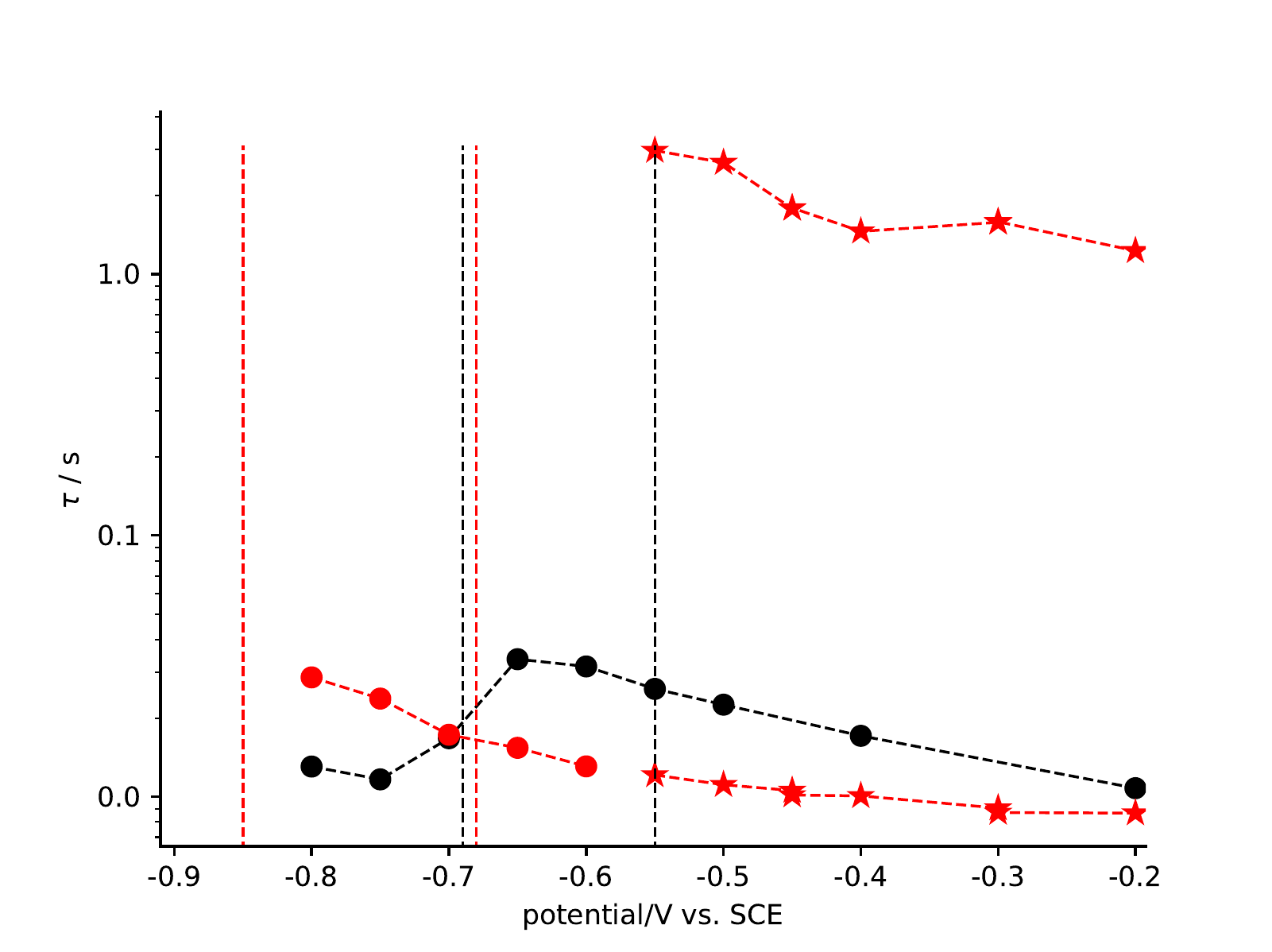}\\
  \includegraphics[height=4.50cm]{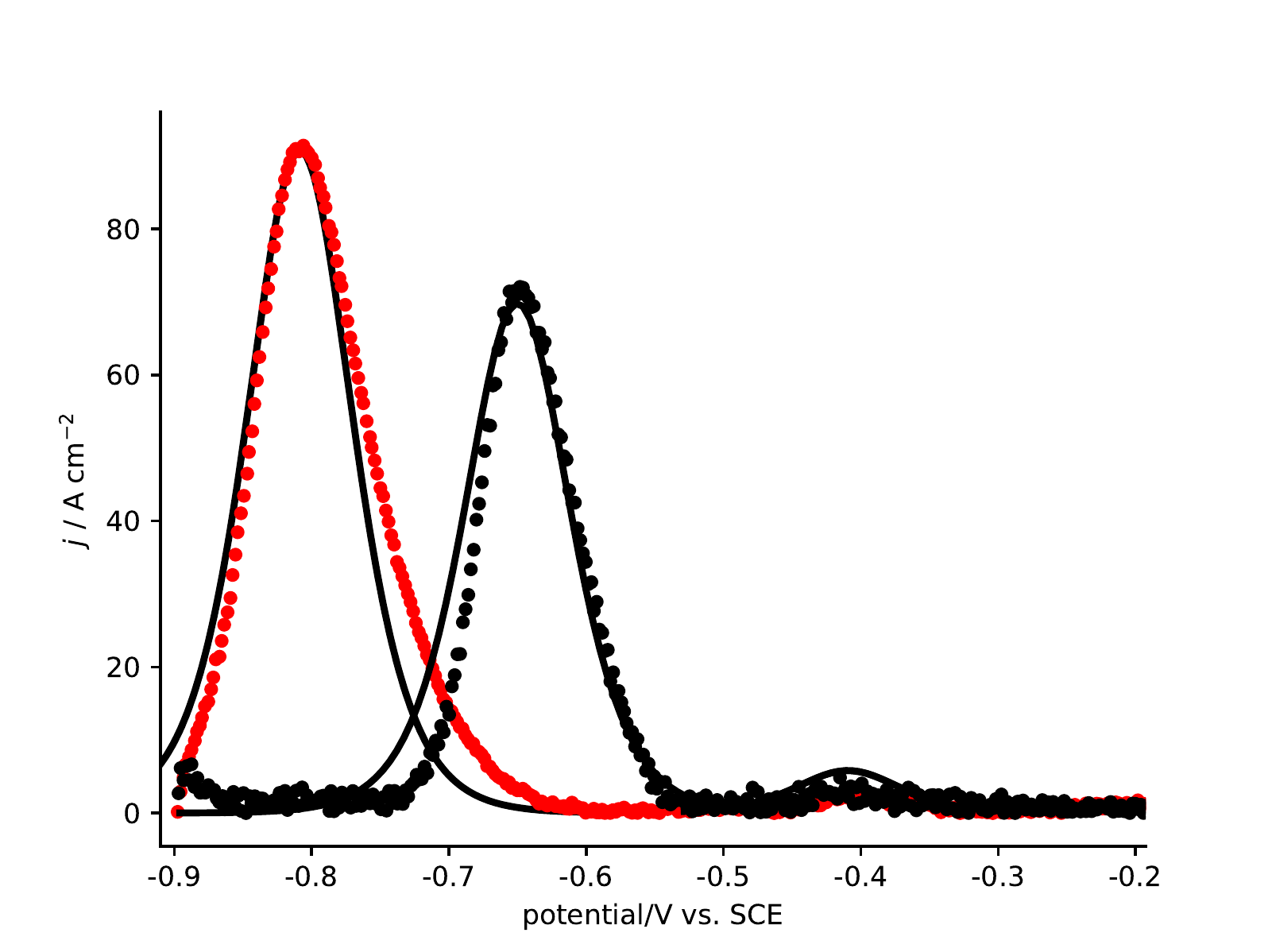}\\
  \caption{Upper panel: Characteristic times of Ag(111) (black) and Ag(100) (red) electrodes in 0.01 M KH$_{2}$PO$_{4}$. 
For Ag(111) (black), the $\tau_{1}$ values were calculated by multiplying the components 
(R$_{1}$$C_{1}$) of $EEC$ in Fig. 1 (c) in the whole potential range. 
For Ag(100) (red), from -0.80 V to -0.60 V the $\tau_{1}$ values were calculated 
from $EEC$ in Fig. 1 (c), and  from -0.55 V to -0.20 V the $\tau_{1}$ and $\tau_{2}$ 
values (R$_{2}$$C_{2}$) were calculated from $EEC$ in Fig. 1 (e). 
Lower panel: Anodic $j-V$ profiles of simulated (black lines) and experimental data (solid circles). 
Scan rate 0.05 V/s.}
\end{figure}

Three potential regions can be distinguished, which have been selected
considering the features observed in the $j-V$ profiles and the relative errors
and $\chi^{2}$ values of the fitting results of the $EIS$ data.
For Ag(111)/KH$_{2}$PO$_{4}$, the lowest relative errors in the
parameters and lowest $\chi^{2}$ values were found to be in the
potential range from -0.70 V to -0.20 V using $EEC$ in Fig. 1 (c) (see Table S4).
This indicates that at these potentials the physical model
that better describes the interface corresponds to a simple surface process,
where the surface coverage of one species can be estimated.

In the case of Ag(100)/KH$_{2}$PO$_{4}$, the results
can be associated to one predominant process
from -0.80 V to -0.60 V ($EEC$ in Fig. 1 (c)),
whereas from -0.55 V to -0.20 V two processes
can be clearly differentiated (red stars obtained
using $EEC$ in Fig. 1 (e): $\tau_{1}$ and $\tau_{2}$),
as can be concluded from the analysis of errors and
$\chi^{2}$ (see Table S5). From these findings it was demonstrated that
the adsorbate surface dynamics is strongly dependent on the
potential applied to the two electrodes.
For $\tau_{1}$ the slowest process is observed from -0.70 to -0.20 V
on the (111) face,
which can be attributed to a higher coordination with Ag atoms of these species.
These findings show
an opposite behavior
when comparing with
other adsorbates on (100) and (111) faces
\cite{doi:10.1002/celc.201800617, doi:10.1021/jacs.8b04903}.

In $j-V$ profiles
two peaks are observed at -0.65 V and
-0.80 V for Ag(111) and Ag(100), respectively (lower panel)
where the solid line is
the simulation of a process that involves only one adsorbed species on
the electrode surface \cite{saveant2006}:

\begin{centering}
\begin{equation}
j = \left( \frac{nF v \Gamma _{0}}{RT \Gamma _{i}}\right)  
\frac{\exp (-nF/RT\Delta \phi)}{(1 + \exp (-nF/RT\Delta \phi))^{2}} \qquad (1) \nonumber
\end{equation}
\end{centering}

where $\Gamma _{0}$ represents a monolayer of the adsorbate, $\Gamma _{i}$
is the surface concentration, and $v$ is the scan rate.
When scanning the potential from negative to positive values,
the following processes can be represented by Eq. (1):

(a) restructuration of the adsorbed layer that
conducts to a different surface geometry, but
always involving the Ag-O bond \cite{doi:10.1021/jacs.8b04903}.

(b) dissociation of H$_{2}$PO$_{4}$$_{(ads)}$
to HPO$_{4}$$_{(ads)}$
\cite{park2017, savizi2011}.
According to these results, we can conclude that
this is a good model to analyze the
Ag(hkl)/H$_{2}$PO$_{4}$$_{(ads)}$
in terms of data collected
during the anodic scan for
the peaks observed between
-0.90 V and -0.50 V.

Bidoia supposed a Langmuir isotherm as an appropriate model for
dihydrogen phosphate adsorption on the platinum surface \cite{bidoia20051}.

Jovi\'c et al.
\cite{jovic1992327} have shown that the adsorption of
NaF, Na$_{2}$SO$_{4}$ and NaCH$_{3}$CO$_{2}$
on Ag(111) occurs with a partial charge transfer at
a measurable rate.
They were the first authors to propose the
inclusion of an additional branch parallel to the double
layer capacitance, i.e. the $EEC$ in Fig 1 (c).
In our case, in the potential region where the peaks are, the $EIS$ results gave the best fit for the $EEC$ in Fig 1 (c).
Integration of $j-V$ peaks at -0.80 V and -0.65 V with respect to the
potential gives 6.7 $\mu$Ccm$^{-2}$ for the (111) surface,
well correlated with the value obtained theoretically \cite{salim2017}.
On the other hand, for the (100) surface the value obtained was 10.4 $\mu$Ccm$^{-2}$,
a bit higher than the previously informed.
Surface charge densities were previously calculated by DFT in the substrate/vacuum
interface for p(4x4) ordered submonolayers of H$_{2}$PO$_{4}$$_{(ads)}$/Ag(hkl)
corresponding to a coverage of 0.065 ML \cite{salim2017}.
Although they were not determined under 
electrochemical conditions, 7.43 $\mu$Ccm$^{-2}$ for Ag(100),
and 7.25 $\mu$Ccm$^{-2}$ 
for Ag(111) were obtained, in good agreement with 
hese experimental results. Table 2 
shows the coverage values obtained as in Fig. 8.
Thus, employing this strategy, our results
indicate that it is indeed possible to obtain a
quantitative evaluation and comparison at the two silver surface facets
of the kinetics of the processes related to H$_{2}$PO$_{4}$$_{(ads)}$ species
(as it can be observed from Fig. 9).

Mostany et al. \cite{mostany20095836}
also reported anodic/cathodic symmetric peaks for
cyclic voltammograms of Pt(111) in 0.1 M HClO$_{4}$ + xM NaH$_{2}$PO$_{4}$
for xM up to 0.01 M. They have reported for 0.01 M NaH$_{2}$PO$_{4}$ a coverage
of 0.22 ML with a surface charge equal to 63 $\mu$cm$^{-2}$.

On Au(111) and Au(100) electrodes, Weber et al. \cite{weber1998}
concluded that the coordination of H$_{2}$PO$_{4}$$_{(ads)}$,
for both electrodes, was two-fold through the two oxygen atoms,
yet they have not informed coverage.

Fig. 10 (upper and middle panel) shows the $\tau_{1}$ and $\tau_{2}$
as a function of potential of Ag(111) and Ag(100) in 0.01 M K$_{3}$PO$_{4}$
between -1.20 V and -0.20 V.
The processes from
-0.85 V to -0.55 V are faster on (100) than
on (111) surface ($\tau_{1}$, middle panel).
The values range
between 3.8 and 9.6 ms.
The interaction that is evidenced from the
voltammograms (Fig. 10 lower panel), i.e., different co-existing adsorption
sites, and the dependence of the adsorption kinetics on the electrode crystal
face and electrode potential are demonstrated.
In the case of $\tau_{2}$, slow processes are involved (from 0.7 s to 7 s),
thus no direct associations can be made with the $j-V$ profiles.

\begin{figure}
\centering
  \includegraphics[height=4.50cm]{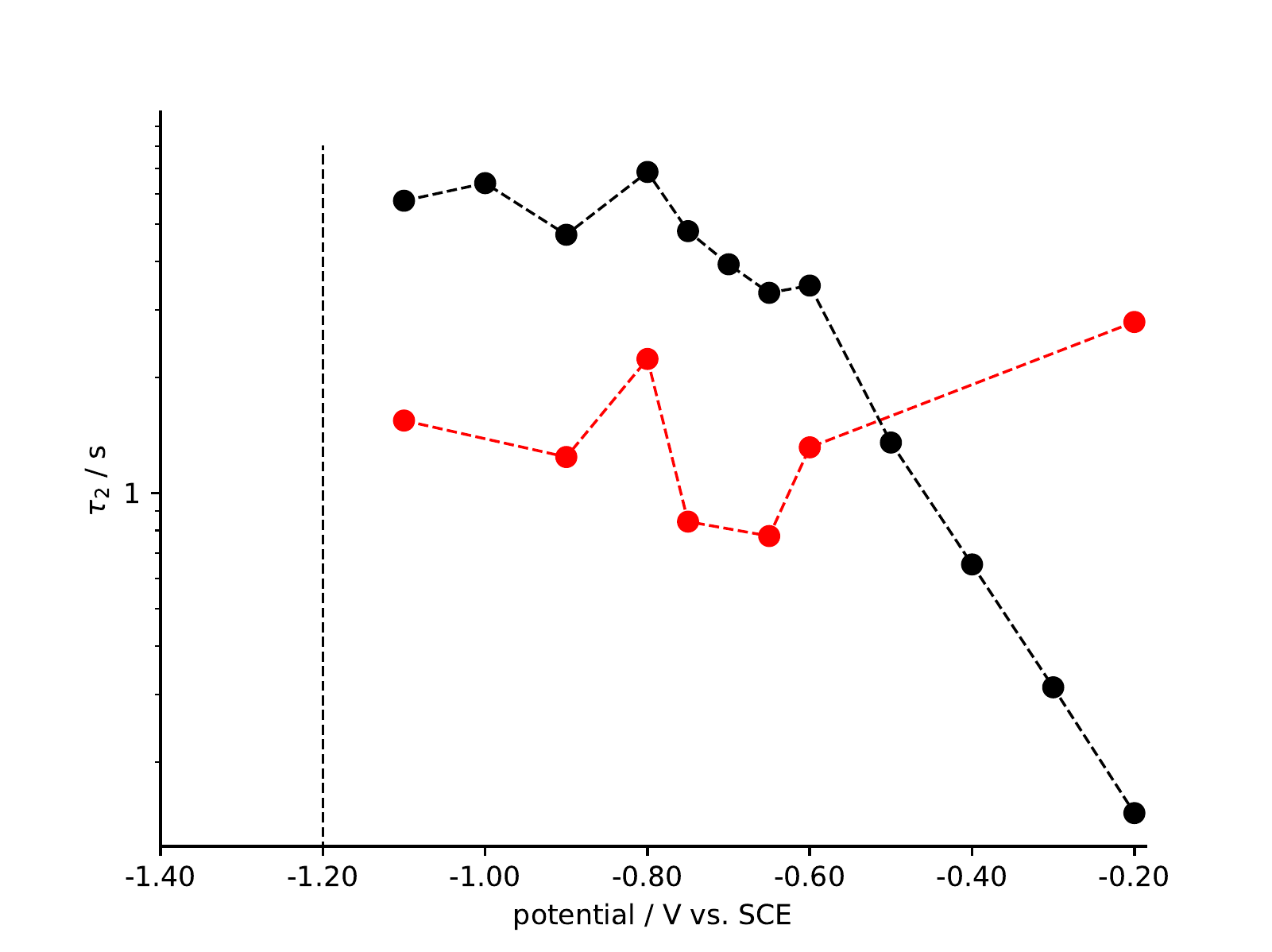}\\
  \includegraphics[height=4.50cm]{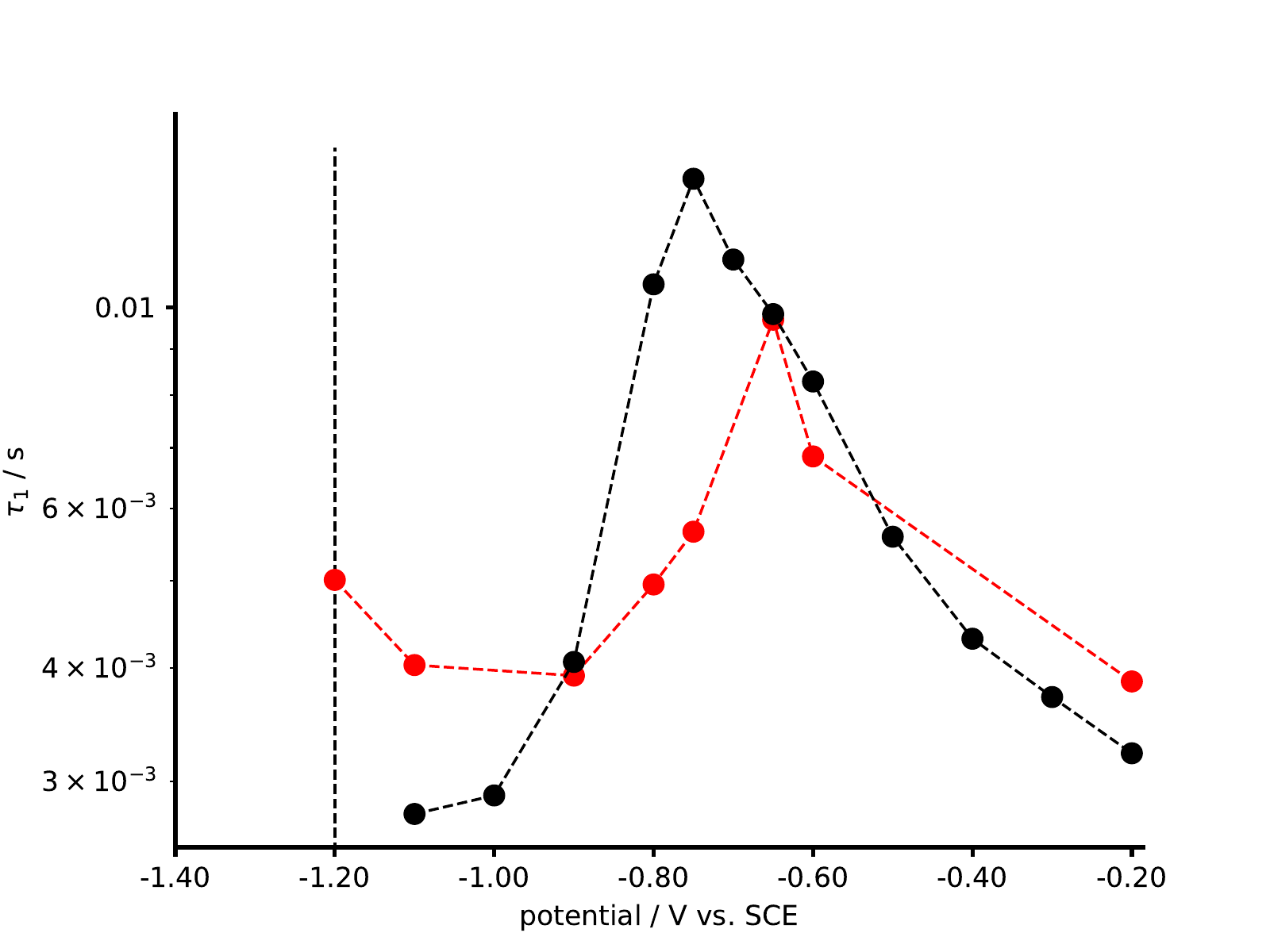}\\
  \includegraphics[height=4.50cm]{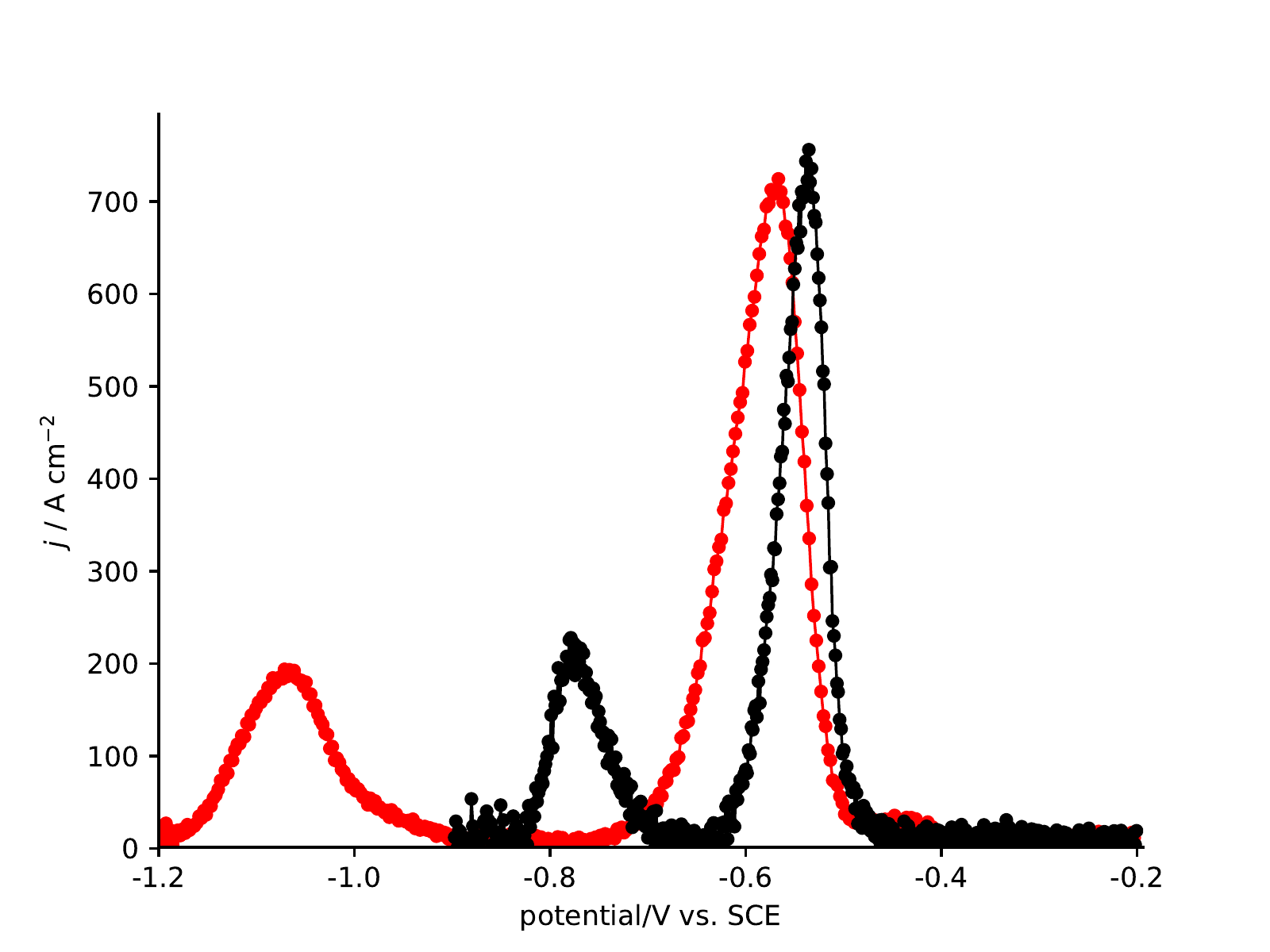}\\
  \caption{Characteristic times: $\tau_{1}$ and $\tau_{2}$ of Ag(111) (black) 
and Ag(100) (red) electrodes in 0.01 M K$_{3}$PO$_{4}$. The $\tau_{1}$ and $\tau_{2}$ 
values were calculated from the $EEC$ in Fig. 1 (e) in the whole potential range. 
Upper panel: $\tau_{1}$ for Ag(111) (black) and Ag(100) (red) electrodes. 
Middle panel: $\tau_{2}$ for Ag(111) (black) and Ag(100) (red) electrodes. 
Lower panel: Anodic $j-V$ profiles of Ag(111) (black) and Ag(100) (red) electrodes. 
Scan rate 0.05 V/s.}
\end{figure}

In the case of Ag(111) surface, the $\tau_{1}$ values range between 2.7 and 14 ms.
The peak at -0.77 V (Fig. 10 lower panel) was identified in Fig. 5 as the contribution
due to co-adsorbed OH$_{(ads)}$. Therefore, according to Ref. \cite{niaura1997surface}
the peak at -0.55 V is related to the restructuring of the phosphate adsorbed species.
As no peaks were observed at potentials negative to -0.77 V when the scan starts at
-1.35 V (see Fig. 5 Upper panel), the coverage could only be calculated
from the chathodic peak at -0.90 V, and is informed in Table 2.
For Ag(100) surface, integration of the anodic peak at -1.10 V in relation to the
potential gives 17.6 $\mu$Ccm$^{-2}$ (see Table S3),
corresponding to a coverage of 0.092 ML,
and the cathodic peak at -1.0 V gave a coverage of 0.058 ML (Table 2).
Phosphate species are widely used as buffer solutions in the study of
important catalytic reactions. Therefore, it is relevant to determine the effect
produced by the phosphate species adlayer on elementary steps, as in $her/hor$ mechanisms
\cite{jackson2019, obata2019, dubouis2017, tian2019, shen2019}.
The present study is the first that reports the study of $EIS$ and $CV$ on Ag(hkl)
in phosphate electrolytes. This information is relevant to the resolution of
multiple-state adsorption of phosphate species on single crystal Ag surfaces and
also to the influence of co-adsorbed anions.

\begin{table}[htb]
\centering
\begin{tabular}{c c c c c c c}
\hline
\shortstack{Electrolyte\\0.01 M} & \shortstack{$\theta$/\\Ag(111)} & \shortstack{$\theta$/\\Ag(111)}  &  \shortstack{$\theta$/\\Ag(100)} &  \shortstack{$\theta$/\\Ag(100)}\\
\hline
 & \shortstack{anodic} & \shortstack{cathodic} & \shortstack{anodic} & \shortstack{cathodic}\\
\hline
H$_{3}$PO$_{4}$     &  0.023 & 0.025 & - & -\\
\hline
KH$_{2}$PO$_{4}$     &  0.030 & 0.033 & 0.054 & 0.058\\
\hline
K$_{3}$PO$_{4}$     & -  & 0.071 & 0.092 & 0.058\\
\hline
\end{tabular}

\caption{Coverage of Ag(111) and Ag(100) surfaces in different electrolytes.
The extent of surface coverage is expressed as the fractional coverage, $\theta$.
Surface atom density of Ag(111) and Ag(100) orientations are
$\sigma$=1.38$\times$10$^{15}$ atoms/cm$^{2}$,
and $\sigma$=1.20$\times$10$^{15}$ atoms/cm$^{2}$, respectively.}
\end{table}

\subsection{Heterogeneous reactions}
In the case of Ag(hkl) in 0.01 M H$_{3}$PO$_{4}$$_{(aq)}$,
the following reactions were postulated:

\begin{centering}
\vspace{0.2cm}
\indent H$_{3}$PO$_{4}$$_{(aq)}$
$\rightarrow$
H$_{2}$PO$_{4}$$^{-}$$_{(aq)}$
+ H$^{+}$$_{(aq)}$  pK$_{a}$ = 2.04 \qquad (2)
\vspace{0.2cm}
\end{centering}\\

and

\begin{centering}
\vspace{0.2cm}
\indent H$_{2}$PO$_{4}$$^{-}$$_{(aq)}$ +
(nH$_{2}$O)$_{(ads)}$/$Ag$(hkl)
$\rightarrow$
((n-x)H$_{2}$O +
H$_{2}$PO$_{4}$)$_{(ads)}$/$Ag$(hkl) +
xH$_{2}$O + e$^{-}$ \qquad (3)
\vspace{0.2cm}
\end{centering}\\

where $Ag$(hkl) is the silver single crystal substrate,
and the subscripts "$(ads)$" and "$(aq)$"
represent the adsorbate and the species in solution,
respectively.

The adsorbed water is displaced by the ion.
Hughes Z.E. et al. have calculated the adsorption
energies of water molecules on both surfaces and
-0.93 eV and -0.84 eV were informed for Ag(100)
and Ag(111), respectively \cite{hughes2013}.
The adsorption energies of H$_{2}$PO$_{4}$$_{(ads)}$ calculated
by us in Ref. [6] were -3.71 eV for Ag(100) and -3.61 eV for Ag(111).
Although anions have greater adsorption energy than water, due to
repulsive lateral interactions between them, an ordered submonolayer
of low coverage is formed according to Eq. (2).

Dorain et al. \cite{dorain1984} have studied the polycrystalline Ag in
different phosphate solutions by SERS.
They observed the Raman vibrational modes of
H$_{3}$PO$_{4}$$_{(aq)}$, and distinction of peaks from
undissociated H$_{3}$PO$_{4}$$_{(ads)}$ was not detected.
This indicates that adsorption did not occur
under their experimental conditions.
However, a weak and characteristic adsorption
for each face orientation takes place under our experimental setup,
as was shown in Fig. 3 and Fig. 8.
According to our calculations \cite{salim2017}
adsorption energies of -0.98 eV (for 100) and -1.08 eV (for 111) were obtained.
These values are in the order of the water adsorption energy
\cite{hughes2013}, thus significant adsorption of this species
is not expected as shown in Ref. \cite{dorain1984}.

Niaura et al. have characterized the structural changes at the interface
when the coordination of the phosphate species to a surface
of a polycrystalline silver electrode moves from tri(bi)- to mono-dentate
as the potential changes towards more negative values \cite{niaura1997surface}.
We have corroborated these findings using silver single crystals \cite{salim2017}.
Therefore, we support the hypothesis sustained through Eq. (2) that for this system
the main adsorbed species is H$_{2}$PO$_{4}$$_{(ads)}$/Ag.
So far, the adsorption of undissociated H$_{3}$PO$_{4}$$_{(ads)}$
has only been detected on Pt(111) at pH = 0.23 by in-situ FTIRS
\cite{willets2019, nart1992}, therefore we have assumed that
Eq. (1) takes place only in solution.

In the case of the KH$_{2}$PO$_{4}$ electrolyte,
where the species of highest concentration are
H$_{2}$PO$_{4}^{-}$ anions,
it has been shown that the main adsorbed species is H$_{2}$PO$_{4}$$_{(ads)}$ \cite{salim2017}.
The H$_{2}$PO$_{4}$$_{(ads)}$/Ag
as the main adsorbate has also been proved for other
systems \cite{savizi2011, salim2017, grunder2019surfaces},
thus we postulate that Eq. (2) also applies to our experimental conditions.

For the adsorption of H$_{2}$PO$_{4}$$_{(ads)}$ on Ag(100),
the most stable configuration is the tri-dentate one,
but this species can also be adsorbed on other configuration.
Thus, the Minimum Energy Path (MEP) at substrate/vacuum interface
corresponding to the change between bi-dentate to tri-dentate configurations
was studied by means of nudged elastic band (NEB) algorithm and DFT calculations,
finding an energy barrier of 0.37 eV \cite{salim2017}.
This is a high barrier for a spontaneous process at room temperature,
but at electrochemical conditions it is expected that all
H$_{2}$PO$_{4}$$_{(ads)}$ species should be in the tri-dentate configuration.

In addition to the vacuum slab method, the double reference model
was used by Savizi et al. \cite{savizi2011}.
They have determined that the main adsorbed species
on Pt(111) was H$_{2}$PO$_{4}$$_{(ads)}$.

In the case of Ag(hkl) in 0.01 M K$_{3}$PO$_{4}$,
the species of higher concentration in the electrolyte are
PO$_{4}^{3-}$ and OH$^{-}$. Thus,
the following reaction
takes place \cite{salim2017}:

\begin{centering}
\vspace{0.2cm}
\indent PO$_{4}$$^{3-}$$_{(aq)}$ +
OH$^{-}$$_{(aq)}$ + (nH$_{2}$O)$_{(ads)}$/$Ag$
$\rightarrow$
((n-x)H$_{2}$O +
OH + PO$_{4}$)$_{(ads)}$/$Ag$ +
xH$_{2}$O + 4 e$^{-}$ \qquad (4)
\vspace{0.2cm}
\end{centering}

For solutions with high OH$^{-}$ concentration,
it has been demonstrated that the presence of surface oxides and
adsorbed hydroxide ions has effective influence on displacing
phosphate adsorbed species on Au and Cu electrodes, effect that
was not observed on Ag surfaces \cite{niaura1997surface}.
For this system we have found competitive adsorption
between OH$^{-}$, HPO$_{4}^{2-}$ and PO$_{4}^{3-}$.

\section{Conclusions}
The following conclusions can be drawn
according to the results discussed above:

- From the dielectric complex permittivity representation of the $EIS$ data,
a semicircle was observed even with the most negative potential values, showing
that $her$ takes place at the same time as the processes associated with the phosphate species.

- We have shown that there is a strong potential dependence of $EIS$ spectra,
and the data can be modeled using the appropriate $EEC$ for each potential range.
At low overpotentials for $her$, the better fit corresponded to the $EEC$ in Fig. 1 (d),
except for the Ag(111)/K$_{3}$PO$_{4}$ system where the lowest errors were given for the $EEC$ in Fig. 1 (f).
At pH 1.60 the R$_{ct}$ was resolved at low overpotentials of $her$ on Ag(100), where
the values were larger than those of the resistance representing the adsorption process.

- In the case of Ag(111)/KH$_{2}$PO$_{4}$, three potential
regions were distinguished in the $j-V$ profiles.
From -1.20 V to -0.90 V, $her$ is taking place.
From -0.85 V to -0.65 V, a stable submonolayer is formed.
From -0.65 V to -0.20 V, two reversible $j-V$ peaks are observed,
that correlate well with the changes observed in the
R$_{1}$ and $\tau_{1}$ values.
For Ag(100)/KH$_{2}$PO$_{4}$ system,
from -0.80 V to -0.60 V one predominant
process was found. From -0.60 V to -0.20 V,
the best fitting was given by the $EEC$
in Fig. 1 (e), which can be explained
by the co-existence of two processes.

- For the Ag(111)/K$_{3}$PO$_{4}$ system, two processes with different kinetics were
identified in terms of $EIS$ fitting results in the same potential window where the
peaks are observed in the voltammograms. The kinetics of adsorption of OH$^{-}$ is
faster (the adsorption resistance decreased) in presence of phosphate species.

- For Ag(111)/H$_{3}$PO$_{4}$ system, a similar coverage of 0.024 ML was calculated
from anodic/cathodic $j-V$ profiles.
Average values of 0.032 ML and 0.056 ML were obtained
for H$_{2}$PO$_{4}$$_{(ads)}$/Ag(111)
and H$_{2}$PO$_{4}$$_{(ads)}$/Ag(100), respectively.
For PO$_{4}$$_{(ads)}$/Ag(111) a value of 0.071 ML was obtained,
well correlated with previously reported theoretical results.
In the case of PO$_{4}$$_{(ads)}$/Ag(100),
values of 0.092 ML and 0.058 ML were obtained
for anodic and cathodic $j-V$ profiles, respectively.

- An adsorbate-induced modification of the
$her$ overpotential was found for all pH.
At pH=1.60 the Ag(100) surface displayed
a relatively higher activity and selectivity
for $her$ as compared with Ag(111).

\section{ACKNOWLEDGMENTS}
The authors thank support from CONICET (Project 15920150100013CO, Res. 2049/16 Argentina)
and SeCyT 33620180100222CB (UNC, C\'ordoba, Argentina).
The authors also wish to thank language assistance by C. Mosconi.

\newpage

\bibliographystyle{unsrt}
\bibliography{ms.bib}

\end{document}


\title{Specific adsorption of phosphate species on Ag (111) and Ag (100) electrodes and their effect at low overpotentials of the hydrogen evolution reaction}
\author{Claudia B. Salim Rosales$^{(1)}$, Mariana I. Rojas$^{(2)}$ and Luc\'ia B. Avalle$^{(1)}$
\thanks{Corrresponding author. e-mail:avalle@famaf.unc.edu.ar}\\
(1) IFEG, CONICET, FaMAF, UNC, C\'ordoba, Argentina.\\
(2) INFIQC (CONICET), Fac. de Ciencias Qu\'imicas (U.N.C.), \\ 
C\'ordoba, Argentina.}

\date{}
\maketitle

\section{$CV$ measurements}
The charges of phosphate species adsorbed on Ag(hkl) electrodes were calculated, after the baseline had been subtracted, from integration of anodic and cathodic curves obtained by cyclic voltammetry between the initial ($\phi_{i}$) and final ($\phi_{f}$) potentials informed in Tables S1-S3.
The $j-V$ profiles for all systems were integrated by the fityk free software (M. Wojdyr, J. Appl. Cryst. 43, 1126-1128 (2010)).

\vspace{1cm}
Table S1

Charges calculated from integration of the $j-V$ profiles for Ag(111) electrode in 0.01 M H$_{3}$PO$_{4}$. The charges are in $\mu$Ccm$^{-2}$. 
Potentials are in V vs. SCE.

\vspace{0.25cm}
\begin{tabular}{c c c c c}
\hline \\
\shortstack{H$_{3}$PO$_{4}$} & \shortstack{charge} & \shortstack{$\phi_{i}$/$\phi_{f}$}  \\
\hline 
anodic     &  5.1 & (-0.74) / (-0.42)  \\
\hline
cathodic    &  5.6 & (-0.74) / (-0.40)  \\
\hline
\end{tabular}

\vspace{1cm}

Table S2

Charges calculated from integration of the $j-V$ profiles for Ag(hkl) electrodes in 0.01 M KH$_{2}$PO$_{4}$. The charges are in $\mu$Ccm$^{-2}$. 
Potentials are in V vs. SCE.

\vspace{0.25cm}
\begin{tabular}{c c c c c}
\hline \\
\shortstack{KH$_{2}$PO$_{4}$\\electrolyte} & \shortstack{charge/\\Ag(111)} & \shortstack{$\phi_{i}$/$\phi_{f}$\\Ag(111)} & \shortstack{charge/\\Ag(100)} & \shortstack{$\phi_{i}$/$\phi_{f}$\\Ag(100)}  \\
\hline 
anodic     &  6.7 & (-0.69) / (-0.55) & 10.4 & (-0.93) / (-0.58)  \\
\hline
cathodic    & 7.4 & (-0.69) / (-0.55) & 11.1 &  (-0.93) / (-0.55)  \\
\hline
\end{tabular}

\vspace{1cm}
Table S3

Charges calculated from integration of the $j-V$ profiles for Ag(hkl) electrodes in 0.01 M K$_{3}$PO$_{4}$. The charges are in $\mu$Ccm$^{-2}$. 
Potentials are in V vs. SCE.

\vspace{0.25cm}
\begin{tabular}{c c c c c}
\hline \\
\shortstack{K$_{3}$PO$_{4}$\\electrolyte} & \shortstack{charge/\\Ag(111)} & \shortstack{$\phi_{i}$/$\phi_{f}$\\Ag(111)} & \shortstack{charge/\\Ag(100)} & \shortstack{$\phi_{i}$/$\phi_{f}$\\Ag(100)}  \\
\hline 
anodic     &  - & - & 17.6 & (-1.24) / (-1.0)  \\
\hline
cathodic    & 15.8 & (-1.07) / (-0.82) & 11.1 &  (-1.11) / (-0.86)  \\
\hline
\end{tabular}

\section{$EIS$ measurements}
The general procedure to fit the $EIS$ data is explained in section 2.3.

Different $EEC$s were fitted for each system in different potential ranges \cite{grden2017, lvovich2012impedance}.
A multistep procedure was addopted:
(1) for the first run
the solution resistance (R$_{s}$)
was fixed to the value of the
high frequency impedance (Z$_{(\infty)}$),
while the rest of the parameters were set free.
(2) for the second run
the parameters obtained from step (1) were reentered
as the new initial values and the fitting was rerun with all parameters free.
(3) this procedure was repeated until the error and $\chi^{2}$ values for a given potential region decreased to the lowest values.

Grde{\'{n}} \cite{grden2017} has studied oxide formation on Ag electrodes in alkaline electrolyte, and has employed a similar procedure and analysis to fit the $EIS$ spectra by the equivalent circuit method.

Fig. S1 and Tables S4-S9 provide an example of the results corresponding to the best fittings, reported as $\chi^{2}$ values and ranges of \%errors of circuit components, for Ag(hkl)/H$_{3}$PO$_{4}$, Ag(hkl)/KH$_{2}$PO$_{4}$ and Ag(hkl)/K$_{3}$PO$_{4}$ systems. A detailed description of Fig. S1 is given in the next section.

\subsection{Selection of $EEC$ in the different potential ranges}
For the selection of the $EEC$s tested in the analysis of $EIS$ data, the results of both the $\chi^{2}$ and \%errors were taken into account, showing strong dependence with potential \cite{morin1996}. The final selection was made considering the physical model that applies to each potential region.

Fig. S1 (upper panel) shows the $\chi^{2}$ values obtained using the $EEC$s of Fig. 1 (c) (blue) and (d) (magenta) for the $EIS$ data fitting of Ag(100)/H$_{3}$PO$_{4}$ system (left). For Ag(111) (right), $EEC$s correspond to (c) (blue) and (e) (magenta).

Fig. S1 (middle panel) shows the $\chi^{2}$ values obtained using the $EEC$s of Fig. 1 (c) (blue) and (e) (magenta) for the $EIS$ data fitting of Ag(hkl)/KH$_{2}$PO$_{4}$ systems.

In this case, the $EIS$ spectra of both surface orientations gave the best fittings with the same $EEC$s.

Fig. S1 (lower panel) shows the $\chi^{2}$ values obtained from the $EIS$ data fitting using $EEC$s of Fig. 1 (e) (blue) and (d) (magenta) for the Ag(100)/K$_{3}$PO$_{4}$ system (left). For the Ag(111)/K$_{3}$PO$_{4}$ system (right), $EEC$s (e) (blue) and (f) (magenta) gave the best fittings.

\subsection{The calculus of $\tau$ values}
The $\tau$s values were calculated by multiplying the R and C elements of the same branch in the electrical equivalent circuit, $\tau_{1}$ = R$_{1}$C$_{1}$ and $\tau_{2}$ = R$_{2}$C$_{2}$.
In the case of Ag(hkl)/KH$_{2}$PO$_{4}$, to calculate the $\tau$ value at -0.90 V for both surface orientations, the (RC) parameters obtained from data fitting using the $EEC$ of Fig. 1 (d) and (f) were used. For the data fitting using the $EEC$ of Fig. 1 (f) at -0.90 V, the $\chi^{2}$ value was 0.1 and the  \%errors of R and C were greater than 50 \%. Therefore, it ended up with the application and the analysis of the $EEC$ of Fig. 1 (d), i.e., the data fitting gave best results in terms of error distribution, low values of both, the $\chi^{2}$ and the \%error (see Tables S4 and S5) for the $EEC$ that considers the charge-transfer process.

In Ag(111)/KH$_{2}$PO$_{4}$ system, from -0.70 V to -0.20 V, $\chi_{}^{2}$ and \% error values corresponding to the data fitting with the $EEC$ of Fig. 1 (c), are lower than the corresponding values obtained from the $EEC$ of Fig. 1 (e).

From -0.85 V to -0.70 V vs. SCE the $\tau_{}$ values calculated from the fitting using the $EEC$ of Fig. 1 (e) are of the same order for both surface orientations, whereas between -0.65 V and -0.20 V
the physical processes ocurring on (111) surface are better described by $EEC$ of Fig. 1 (c), which presented the lowest errors in this potential range.

In Ag(hkl)/K$_{3}$PO$_{4}$ systems, the R$_{2}$ values calculated using the $EEC$ (e) were lower than R$_{1}$.

In this case, the $EEC$s of Fig. 1 (e) and (f) can be applied for a surface coverage containing two adsorbates with different physical/chemical properties.

From -0.80 V to -0.20 V, as observed for the data fitting using the $EEC$ of Fig. 1 (e) for Ag(111)/K$_{3}$PO$_{4}$ system,
it was possible to distinguish two processes from the analysis of the parameters.

\vspace{1cm}

Table S4. R$_{1}$ and R$_{2}$ relative errors and $\chi^{2}$ calculated from the data fitting for $EIS$ employing the $EEC$s of Fig. 1 for Ag(111)/KH$_{2}$PO$_{4}$ system in different potential ranges.

For the $EEC$ in Fig. 1 (c), the \% errors in C$_{1}$ values ranged between 0.2 and 14 \%. For the $EEC$ in Fig. 1 (e), C$_{1}$ values were from 0.5 to 17\%, and C$_{2}$ values from 14\% to 40\%. As there was no charge transfer at potentials more positive than -0.90 V, the $EEC$ in Fig. 1 (d) was used only at -0.90 V.

\vspace{0.1cm}

\begin{tabular}{c c |c c}

$EEC$  \qquad \qquad  & $\chi^{2}$ & \shortstack{relative errors of\\the circuit\\components/\%}  \\ 

\hline
 & &  R$_{1}$ & R$_{2}$  \\ \cline{3-4}
a      & $>$ 1 & $>$ 50 & -\\
\hline
b      & $>$ 1 & $>$ 50 & - \\
\hline
       & $>$ 1 & 7.0-10.7  & -  \\
       & (from -0.9 V to -0.85 V)   \\ \cline{2-4}
c       & 0.1 (-0.80 V) & 1.4-2.9  & -  \\ \cline{2-4}
       & 0.01-0.003 & 0.2-2.7    \\
       & (from -0.7 V to -0.2 V) & & -  \\ 
\hline
d       & 0.02 & 1.1  & 0.7   \\ 
        & (-0.90 V) & & \\
\hline
       & 0.01-0.009 & 0.7-16.3 & 0.42-0.53  \\
e       & (from -0.9 V to -0.75 V) &  &   \\ \cline{2-4}
       & 0.02-0.009 & 24.8-48.9  & 1.7-17.4 \\
       & (from -0.7 V to -0.2 V) &  &  \\ 
\hline
\end{tabular} \\
\vspace{0.5cm}

Table S5. R$_{1}$ and R$_{2}$ relative errors and $\chi^{2}$ calculated from the data fitting employing the $EEC$ of Fig. 1 for Ag(100)/KH$_{2}$PO$_{4}$ system in different potential ranges.

From -0.80 V to -0.55 V for the $EEC$ in Fig. 1 (c), \% errors in C$_{1}$ values ranged between 0.7 and 1.7\%. From -0.55 V to -0.20 V, the $EEC$ in Fig. 1 (e) gave \% errors in C$_{1}$ values between 7.6 and 19\%, and for C$_{2}$ values from 1.2 to 18\%. As there was no charge transfer at potentials more positive than -0.90 V, the $EEC$ in Fig. 1 (d) was used only at -0.90 V.
\vspace{0.1cm}

\begin{tabular}{c c |c c}

$EEC$  \qquad \qquad  & $\chi^{2}$ & \shortstack{relative errors of\\the circuit\\components/\%}  \\ 
\hline
 & &  R$_{1}$ & R$_{2}$  \\ \cline{3-4}
a      & $>$ 1 & $>$ 50 & -\\
\hline
b      & $>$ 1 & $>$ 50 & - \\
\hline
       & $>$ 1  & $>$ 50  & -  \\
       & (-0.90 V)   \\ \cline{2-4}
       & 0.4-0.02 & 0.6-2.4  & -    \\
c       & (from -0.80 V to -0.6 V) & 1.4-2.9  & -    \\ \cline{2-4}
       & $>$ 1 & & - \\ 
       & \shortstack{(from-0.50 V to -0.30 V)} & $>$ 50  &     \\ \cline{2-4} 
       & 0.06 & 1.43  & -   \\
       & (-0.20 V) &   &     \\ \cline{2-4}
\hline
d       & 0.02 & 1.0  & 1.1   \\ 
        & (-0.90 V) & & \\
\hline
       & 0.1-0.04 & 1.2-18.9 & 1.0-18.9  \\
       & (from -0.9 V to -0.20 V) & & \\
e       &  &  & $>$ 50   \\ 
        &  &  & \shortstack{(from -0.90 V\\to -0.85 V)}   \\ \cline{2-4}
\hline
\end{tabular} \\
\vspace{0.5cm}

\vspace{0.5cm}

Table S6. R$_{1}$ and R$_{2}$ relative errors and $\chi^{2}$ calculated from the data fitting employing the $EEC$ of Fig. 1 for Ag(100)/H$_{3}$PO$_{4}$ system in different potential ranges.

The \% errors in C$_{1}$ values ranged between 0.7 and 1.8\% for the $EEC$ in Fig. 1 (c). The $EEC$s in Fig. 1 (e) and (f) gave \% errors in C$_{1}$ values between 20 and 78\%, and for C$_{2}$ values from 16 and 70\%. The $EEC$ in Fig. 1 (d) was used from -0.70 V to -0.6 V and \% errors in C$_{1}$ values were between 0.3 and 0.9\%.
\vspace{0.1cm}

\begin{tabular}{c c |c c}

$EEC$  \qquad \qquad  & $\chi^{2}$ & \shortstack{relative errors of\\the circuit\\components/\%}  \\ 

\hline
 & &  R$_{1}$ & R$_{2}$  \\ \cline{3-4}
a      & $>$ 1 & $>$ 50 & -\\
\hline
b      & $>$ 1 & $>$ 50 & - \\
\hline
       & 0.4-0.03 & 4.8-0.8  & -    \\
c       & \shortstack{( -0.70 V and\\from -0.6 to -0.20 V)} &   &  \\ \cline{2-4}
       & $>$ 1 & 6.3  &  \\ 
       & \shortstack{(-0.675 V)} &   & -    \\ \cline{2-4} 
\hline
       & 0.01-0.1 & 0.3-1.3 & -  \\
d       & (from -0.70 V to -0.60 V) & & \\
\hline
e       & 0.2 & $>$ 20  & 1.5-1.6   \\ 
       & (from -0.40 V to -0.20 V) & & \\
\hline
f       & $>$ 1 & $>$ 50  & $>$ 50   \\ 
\hline
\end{tabular} \\
\vspace{0.5cm}

\vspace{0.5cm}

Table S7. R$_{1}$ and R$_{2}$ relative errors and $\chi^{2}$ calculated from the data fitting employing the $EEC$s of Fig. 1 for Ag(111)/H$_{3}$PO$_{4}$ system in different potential ranges.

The variability in the \%errors of C$_{1}$ and C$_{2}$ values was between 1.3 and 3.3\% for the different circuits, except for the $EEC$ of Fig. 1 (f) where \% errors were $>$ 20.
\vspace{0.1cm}

\begin{tabular}{c c |c c}

$EEC$  \qquad \qquad  & $\chi^{2}$ & \shortstack{relative errors of\\the circuit\\components/\%}  \\ 

\hline
 & &  R$_{1}$ & R$_{2}$  \\ \cline{3-4}
a      & $>$ 1 & $>$ 50 & -\\
\hline
b      & $>$ 1 & $>$ 50 & - \\
\hline
       & $>$ 1 & $>$ 20  & -    \\
c       & \shortstack{( -0.675 V)} &   &  \\ \cline{2-4}
       & 0.2-0.3 & 1.4-1.7  & - \\ 
       & \shortstack{(from -0.625 V to -0.20 V)} &   &    \\ 
\hline
       & 0.09 & 0.8-1.1 & 7.3  \\
       & (-0.675 V) & & \\  \cline{2-4}
d       & 0.04-0.1 & 0.8-1.1 & $>$ 20  \\
       & (from -0.625 V to -0.20 V) & & \\
\hline
       & 0.04-0.1 & 0.8-1.0  & 7.0-15.8     \\ 
e        & \shortstack{(from -0.675 V to -0.625 V)} & & \\ \cline{2-4}
        & 0.04-0.1 & 0.8-1.3  & $>$ 20 \\
        & \shortstack{(from -0.60 V to -0.20 V)} & & \\
\hline
f        & 0.03-0.1 & $>$ 20  &  1.4-7.1  \\ 
        & \shortstack{(from -0.675 V to -0.20 V)} & & \\ 
\hline

\end{tabular} 

\vspace{1cm}
At -0.675 V for Ag(hkl)/H$_{3}$PO$_{4}$ systems the $EEC$ of Fig. 1 (c) gave bad fitting results. From -0.60 V to -0.20 V good fitting results were obtained for both surfaces. The $EEC$ of Fig. 1 (e) could not be fitted from -0.70 V to -0.50 V for the Ag(100)/H$_{3}$PO$_{4}$ system, and from -0.40 V to -0.20 V even if low values of $\chi^{2}$ were obtained, the error in the R$_{1}$ was $>$ 20 \%.
Instead for Ag(111)/H$_{3}$PO$_{4}$ system, the fitting results from the $EEC$ of Fig. 1 (e) were good from -0.675 V to -0.625 V, and from -0.60 V to -0.20 V the $\chi^{2}$ and errors were found to be equal to the (100) surface, i.e.,
low $\chi^{2}$ but R$_{1}$ parameter errors $>$ 20 \% were found.

\vspace{1cm}

Table S8.
R$_{1}$ and R$_{2}$ relative errors and $\chi^{2}$ calculated from the data fitting employing the $EEC$s in Fig. 1 for Ag(100)/K$_{3}$PO$_{4}$ system in different potential ranges.

The variability in the \%error of C$_{1}$ and C$_{2}$ values was between 4.1 and 6.7 \% for the $EEC$s (c) and (d). In the case of C$_{1}$ obtained by the $EEC$ (e), the \% error only at -1.38 V was 78\%, for the other potentials the error was less than 3.8\%. The \% errors for C$_{2}$ (f) from -1.20 V to -1.40 V were no lower than 58\%.
\vspace{0.1cm}

\begin{tabular}{c c |c c}

$EEC$  \qquad \qquad  & $\chi^{2}$ & \shortstack{relative errors of\\the circuit\\components/\%}  \\ 

\hline
 & &  R$_{1}$ & R$_{2}$  \\ \cline{3-4}
a      & $>$ 1 & $>$ 50 & -\\
\hline
b      & $>$ 1 & $>$ 50 & - \\
\hline
       & $>$ 1 & 13-20  & -    \\
c       & \shortstack{(from -1.38 V to -1.10 V)} &   &  \\ \cline{2-4}
       & 0.4-1.0 & 9.5-14.9  & - \\ 
       & \shortstack{(from -0.90 V to -0.20 V)} &   &    \\  
\hline
       & 0.005-0.5 & 0.6-7.8 & 0.4-12.1  \\
       & (from-1.38 V to -1.10 V) & & \\  \cline{2-4}
d       & 0.1-0.5 & 2.8-7.8 & 12-16.1  \\
       & (from -0.75 V to -0.20 V) & & \\
\hline
e       & 0.03-0.2 & 1.7-16.9  & 1.5-9.6     \\ 
        & \shortstack{(from -1.25 V to -0.20 V)}  & \\
\hline
f        & 0.001-0.05 & 2.2-5.1  &  0.3-1.9   \\ 
        & \shortstack{(from -1.38 V to -1.20 V)} & & \\ 
\hline
\end{tabular} 

\vspace{1cm}


Table S9. R$_{1}$ and R$_{2}$ relative errors and $\chi^{2}$ calculated from the data fitting employing the $EEC$s of Fig. 1 for Ag(111)/K$_{3}$PO$_{4}$ system in different potential ranges.

The variability in the \%error of C$_{1}$ values was between 1.3 and 3.3\%  for (c) and (d) circuits. For (e) and (f) from -1.20 V to -1.40 V their errors were no lower than 47\%, whereas from -0.55 V to -0.20 V their values were between 1 and 19\%.
\vspace{0.1cm}

\begin{tabular}{c c |c c}

$EEC$  \qquad \qquad  & $\chi^{2}$ & \shortstack{relative errors of\\the circuit\\components/\%}  \\ 

\hline
 & &  R$_{1}$ & R$_{2}$  \\ \cline{3-4}
a      & $>$ 1 & $>$ 50 & -\\
\hline
b      & $>$ 1 & $>$ 50 & - \\
\hline
       & 0.5 & 8.2  & - \\ 
       & \shortstack{(-1.30 V)} &    &    \\ \cline{2-4}  
c       & $>$ 1 & $>$ 20  & -    \\
       & \shortstack{(from -1.10 V and -1.00 V)} &   &  \\ \cline{2-4}
       & 0.4-0.5  & $>$ 20  & - \\ 
       & \shortstack{(from -0.80 V to -0.40 V)} &    &    \\ \cline{2-4}  
\hline
       & 0.01-0.2 & 0.9-4.5 & 1.0-8.3  \\
d       & (from -1.30 V to -0.750) & & \\  
\hline
       & 0.01-0.07 & 2.5-13.6  & 0.8-2.2    \\ 
e        & \shortstack{(from -1.3 V to -0.60 V)}  & \\ \cline{2-4}
        & 0.03-0.07 & 13.6-33.5  & 1.5-2.7 \\
        & \shortstack{(from -0.50 V to -0.20 V)} & & \\
\hline
f        & 0.001-0.03 & 19.0-22.0  &  0.9-18.3  \\ 
        & \shortstack{(from -1.30 V to -1.20 V)} & & \\ 
\hline
\end{tabular} 

\section{Additional Tables showing the resistance and capacitance parameters with their corresponding $\chi^{2}$ and \% errors} 

 Table S10.
R$_{ct}$, R$_{1}$ and C$_{1}$ values calculated from $EEC$ of Fig. 1 (d) for Ag(100)/H$_{3}$PO$_{4}$ system from -0.675 V to -0.60 V vs. SCE.

\vspace{0.1cm}

\begin{tabular}{c c c c c c c c}

\shortstack{potential/\\V}  & \shortstack{R$_{ct}$/\\$\Omega$cm$^{2}$} & \shortstack{error$_{R_{ct}}$/\\ \%} & \shortstack{R$_{1}$/\\$\Omega$cm$^{2}$} & \shortstack{error$_{R_{1}}$/\\ \%} & \shortstack{C$_{1}$/\\$\mu$Ccm$^{-2}$} & \shortstack{error$_{C_{1}}$/\\ \%} & $\chi^{2}$  \\
\hline 
-0.675      & 23731.2 & 1.4  & 170.6 & 0.3 & 15.2 & 0.3 & 0.004\\
-0.650      & 35639.3 & 1.8 & 170.0 & 0.3 & 13.7 & 0.3 & 0.004 \\
-0.625      & 49382.9 & 2.3 & 169.7 & 0.4 & 12.6 & 0.3 & 0.1 \\
-0.600      & 56606.2 & 6.2 & 170.3 & 0.9 & 11.7 & 0.9 & 0.04 \\
\end{tabular} \\
\vspace{0.5cm}

For all systems, the fitting results showed that the values corresponding to R$_{ct}$ in the $EEC$s of Fig. 1 (d) and (f) were always higher than those of the resistances in the (R$_{1}$C$_{1}$) and (R$_{2}$C$_{2}$) branches.

\vspace{1cm}
Table S11.

R$_{1}$ values obtained from $EIS$ data fitting using the $EEC$ in Fig. 1 (c) for Ag(111)/KOH system from -0.80 V to -0.2 V vs. SCE. The greater changes were seen in the resistance values, whereas variability in C$_{1}$ values was within 10 \%.

\vspace{0.1cm}

\begin{tabular}{c c c c}
\qquad &   Ag(111)/KOH &  &    \\
potential / V \qquad \qquad  & R$_{1}$ / $\Omega$cm$^{2}$ & error$_{R_{1}}$ / \% & $\chi^{2}$  \\
\hline
-0.80      & 0.137 & 5.0  & 0.3 \\
-0.75      & 0.139 & 4.9 & 0.3  \\
-0.70      & 0.140 & 3.0 & 0.1  \\
-0.65      & 0.143 & 2.2 & 0.3  \\
-0.60      & 0.145 & 12.7 & 0.3 \\
-0.50      & 0.146 & 3.4 & 0.1  \\
-0.40      & 0.146 & 2.5 & 0.09 \\
-0.30      & 0.148  & 2.0 & 0.05 \\
-0.20      & 0.156  & 7.6 & 0.03 \\
\end{tabular} \\
\vspace{0.5cm}

Table S12

Results of $EIS$ data fitting with the $EEC$ of Fig. 1 (e) for Ag(100)/KH$_{2}$PO$_{4}$ system.
From -0.55 V to -0.20 V, the addition of new components to the circuit (e) compared to (c) improved the goodness of the fit (lowering of $\chi^{2}$). Distribution of errors was also improved. The fitting results are reported as $\chi^{2}$ values and porcentual errors for each parameter.
\vspace{0.5cm}

\begin{tabular}{c c c c c c c}
\qquad &  & Ag(100)/KH$_{2}$PO$_{4}$ &  &  &  & \\
potential / V \qquad \qquad  & \shortstack{R$_{1}$\\$\Omega$cm$^{2}$} & \shortstack{error$_{R_{1}}$\\ \%} & \shortstack{R$_{2}$\\ $\Omega$cm$^{2}$} & \shortstack{error$_{R_{2}}$\\ \%} & $\chi^{2}$ \\
\hline
-0.55      & 52974.2 & 7.2 & 228.3 & 1.0 & 0.09 \\
-0.50      & 64130.0 & 7.7 & 232.1 & 0.9 & 0.08  \\
-0.45      & 62026.2 & 8.6 & 225.8 & 1.7 & 0.09 \\
-0.40      & 58034.2 & 10.3 & 237.7 & 1.1 & 0.08 \\
-0.30      & 77542.6  & 8.9 & 230.1 & 1.6 & 0.08 \\
-0.20      & 84305.7  & 12.5 & 241.8 & 1.0 & 0.09 \\
\end{tabular} \\
\vspace{0.5cm}

Table S13

Results of $EIS$ data fitting with the $EEC$ of Fig. 1 (e) for Ag(111)/K$_{3}$PO$_{4}$ system.
The addition of new components to the circuit (e) compared to (c) improved the goodness of the fit (lowering of $\chi^{2}$). The fitting results are reported as $\chi^{2}$ values and \% errors for each parameter.
\vspace{0.5cm}

\begin{tabular}{c c c c c c c}
 &   & Ag(111)/K$_{3}$PO$_{4}$  &  & &  & \\
\shortstack{potential/\\V}   & \shortstack{R$_{1}$/\\$\Omega$cm$^{2}$}  &  \% & \shortstack{R$_{2}$/\\$\Omega$cm$^{2}$} & \% & $\chi^{2}$  \\
\hline
-1.30      & 3300.3 & 2.5  & 92.5 & 2.2 & 0.04  \\
-1.10      & 24237.5 & 8.4 & 91.3 & 3.8 & 0.07 \\
-1.00      & 31112.6 & 6.7 & 91.1 & 2.0 & 0.07 \\
-0.80      & 39735.7 & 8.8 & 90.4 & 1.0 & 0.01 \\
-0.75      & 39691.3 & 9.4 & 91.0 & 0.8 & 0.01  \\
-0.70      & 38846.5 & 10.3 & 91.6 & 1.0 & 0.01\\
-0.65      & 37318.8 & 11.47 & 92.2 & 1.2 & 0.02\\
-0.60      & 46742.4 & 13.55 & 93.4 & 1.5 & 0.03\\
-0.50      & 62491.2 & 21.9 & 94.9 & 1.6 & 0.03 \\
-0.40      & 48837.9 & 27.9 & 96.1 & 1.9 & 0.04 \\
-0.30      & 25071.3 & 33.5 & 97.1 & 2.5 & 0.06 \\
-0.20      & 11485.1 & 31.9 & 98.0 & 2.7 & 0.07 \\
\end{tabular}

\newpage

\begin{figure}
\centering
  \includegraphics[height=5.50cm]{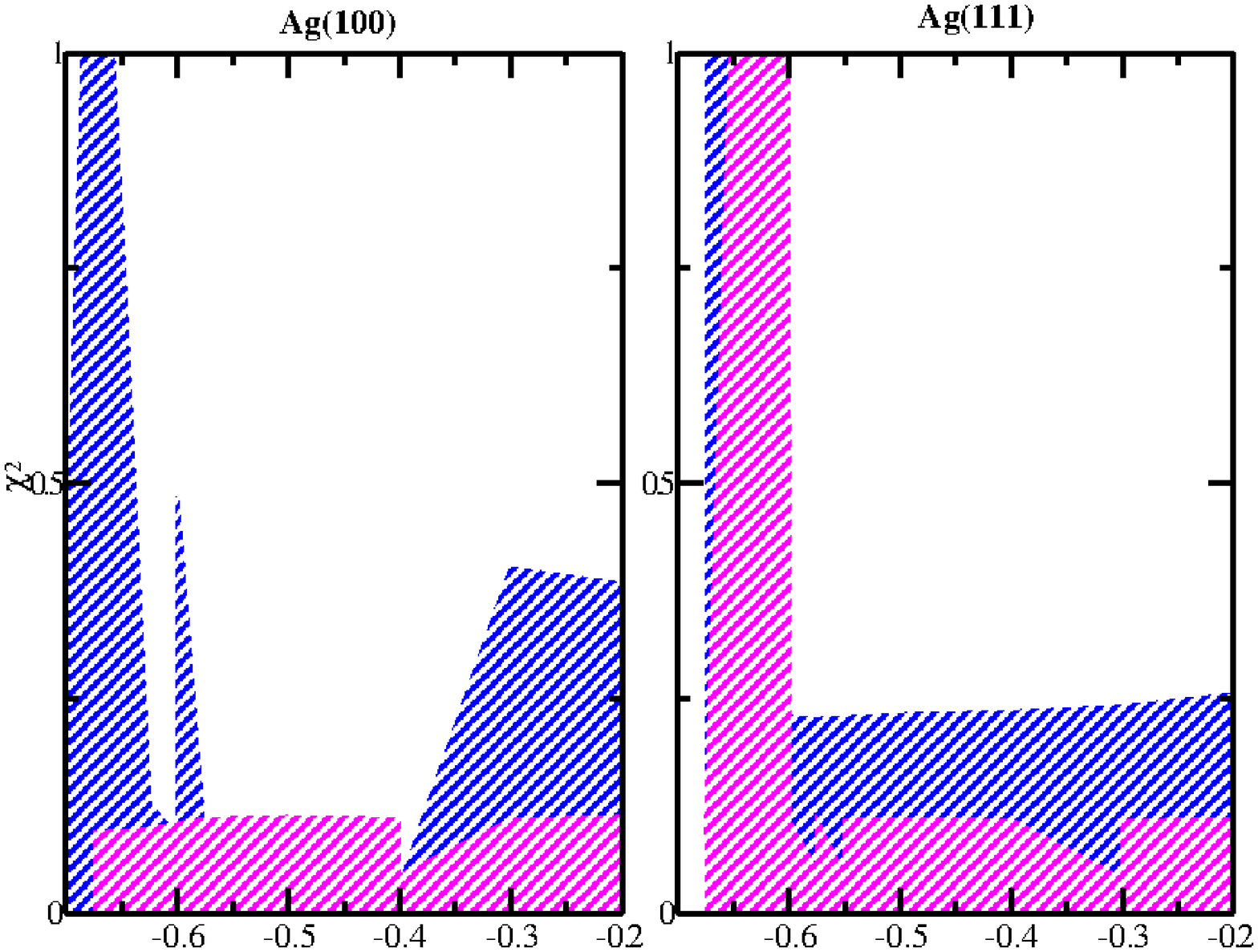} \\
  \includegraphics[height=5.50cm]{figureSupplement1b.eps} \\
  \includegraphics[height=5.50cm]{figureSupplement1c.eps} \\
  \caption{$\chi^{2}$ as a function of potential of Ag(111) and Ag(100) electrodes in 0.01 M electrolytes. Upper panel: For the Ag(100)/H$_{3}$PO$_{4}$ system (left) the fitting corresponds to $EEC$ (c) (blue) and $EEC$ (d) magenta, and for the Ag(111)/H$_{3}$PO$_{4}$ system (right) the fitting corresponds to $EEC$ (c) (blue) and $EEC$ (e) magenta. Middle panel: For the Ag(hkl)/KH$_{2}$PO$_{4}$ systems, $\chi^{2}$ values derived from the $EIS$ data fitting using $EEC$s of Fig. 1 (c) (blue) and (e) magenta. Lower panel: For the Ag(100)/K$_{3}$PO$_{4}$ system (left), $\chi^{2}$ values result from the $EIS$ data fitting by the $EEC$s of Fig. 1 (e) (blue) and (d) (magenta). For the Ag(111)/K$_{3}$PO$_{4}$ system (right), (e) (blue) and (f) (magenta).}
\end{figure}

\bibliographystyle{plain}
\bibliography{supplement}